\documentclass{emulateapj}

\usepackage{color}
\usepackage[usenames,dvipsnames,svgnames,table]{xcolor}
\usepackage{subfigure}
 \usepackage{graphicx}
 \usepackage{threeparttable}
\usepackage{rotating}
\usepackage{lipsum}
\usepackage{arydshln}
\usepackage{amsmath}
\usepackage{hyperref}
\usepackage{booktabs}
\usepackage{xcolor}
\usepackage{soul}

\newcommand{\jerzap}[1]{}

\begin{document}

\title{Spectrophotometric Redshifts in the Faint Infrared Grism Survey: Finding Overdensities of Faint Galaxies}

\author{John Pharo\altaffilmark{1},
Sangeeta Malhotra\altaffilmark{1},
James Rhoads\altaffilmark{1},
Russell Ryan \altaffilmark{2},
Vithal Tilvi\altaffilmark{1},
Norbert Pirzkal\altaffilmark{2},
Steven Finkelstein\altaffilmark{3},
Rogier Windhorst\altaffilmark{1},
Norman Grogin\altaffilmark{2},
Anton Koekemoer\altaffilmark{2},
Zhenya Zheng\altaffilmark{4},
Nimish Hathi\altaffilmark{2},
Keunho Kim\altaffilmark{1},
Bhavin Joshi\altaffilmark{1},
Huan Yang\altaffilmark{1},
Lise Christensen\altaffilmark{5},
Andrea Cimatti\altaffilmark{6,7},
Jonathan P. Gardner\altaffilmark{8},
Nadia Zakamska\altaffilmark{9},
Ignacio Ferreras\altaffilmark{10},
Pascale Hibon\altaffilmark{11}, and
Anna Pasquali\altaffilmark{12}}

\altaffiltext{1}{School of Earth \& Space Exploration, Arizona State University, Tempe, AZ. 85287-1404, USA}
\altaffiltext{2}{Space Telescope Science Institute, Baltimore, MD. 21218, USA}  
\altaffiltext{3}{Department of Astronomy, The University of Texas at Austin, Austin, TX. 78712, USA}
\altaffiltext{4}{Shanghai Astronomical Observatory, Chinese Academy of Sciences, Shanghai 200030, China}
\altaffiltext{5}{1 Dark Cosmology Centre, Niels Bohr Institute, University of Copenhagen, Juliane Maries Vej 30, 2100 Copenhagen, Denmark  }
\altaffiltext{6}{Department of Physics and Astronomy (DIFA), University of Bologna, Via Gobetti 93/2, I-40129, Bologna, Italy }
\altaffiltext{7}{INAF - Osservatorio Astrofisico di Arcetri, Largo E. Fermi 5, I-50125, Firenze, Italy}
\altaffiltext{8}{Astrophysics Science Division, Goddard Space Flight Center, Code 665, Greenbelt, MD 20771, USA }
\altaffiltext{9}{Department of Physics \& Astronomy, Johns Hopkins University, Bloomberg Center, 3400 N. Charles St., Baltimore, MD 21218, USA}
\altaffiltext{10}{Mullard Space Science Laboratory, University College London, Holmbury St Mary, Dorking, Surrey RH5 6NT, UK}
\altaffiltext{11}{ESO, Alonso de Cordova 3107, Santiago, Chile}
\altaffiltext{12}{Astronomisches Rechen-Institut, Zentrum fuer Astronomie, Universitaet Heidelberg, Moenchhofstrasse 12 - 14, D-69120 Heidelberg, Germany}



\begin{abstract}
  We improve the accuracy of photometric redshifts by including low-resolution spectral data from the G102 grism on the Hubble Space Telescope, which assists in redshift determination by further constraining the shape of the broadband Spectral Energy Disribution (SED) and identifying spectral features. The photometry used in the redshift fits includes near-IR photometry from FIGS+CANDELS, as well as optical data from ground-based surveys and HST ACS, and mid-IR data from Spitzer. We calculated the redshifts through the comparison of measured photometry with template galaxy models, using the EAZY photometric redshift code. For objects with F105W $< 26.5$ AB mag with a redshift range of $0 < z < 6$, we find a typical error of $\Delta z = 0.03 * (1+z)$ for the purely photometric redshifts; with the addition of FIGS spectra, these become $\Delta z = 0.02 * (1+z)$, an improvement of 50\%.  Addition of grism data also reduces the outlier rate from 8\%  to 7\% across all fields. With the more-accurate spectrophotometric redshifts (SPZs), we searched the FIGS fields for galaxy overdensities. We identified 24 overdensities across the 4 fields. The strongest overdensity, matching a spectroscopically identified cluster at $z=0.85$, has 28 potential member galaxies, of which 8 have previous spectroscopic confirmation, and features a corresponding X-ray signal. Another corresponding to a cluster at $z=1.84$ has 22 members, 18 of which are spectroscopically confirmed. Additionally, we find 4 overdensities that are detected at an equal or higher significance in at least one metric to the two confirmed clusters.
\end{abstract}


\section{Introduction}

The redshift interval $z \sim 1 - 3$ includes the era of peak star formation, and it hosts the greatest density of galaxy mergers \citep{md14}. Measuring accurate redshifts of such distant galaxies is a difficult problem, and many objects in this range lack the high-resolution, ground-based spectroscopy that produces the most-accurate redshifts, particularly the less-massive, fainter objects. The method of fitting spectrophotometric grism redshifts provides the opportunity to get more precise redshift measurements for such objects.

Observing the objects present in this cosmic epoch provides vital information in the study of the formation and assembly of galaxies and of large scale structure (LSS) in the transition from the epoch of reionization to the modern low-redshift universe, and is vital to understanding our cosmic origins. However, the study of objects at such substantial redshifts necessarily introduces completeness problems: as the distance increases, lower-luminosity objects become more difficult to measure at a useful signal level, and thus may be rendered indistinguishable from low-redshift objects that are particularly faint or dust-extinguished. When conducting a study of high-redshift objects, the loss of the faint population biases the sample, and limits the conclusions that may be drawn \citep{bou15, fink15}.

Mitigating this issue requires deep observations of the faint galaxy population to address the problem of completeness, and in order to be useful, those observations will require definitive measurements of that population's redshift. Determination of a galaxy's redshift generally requires measurement of easily identifiable features in the galaxy's spectrum, such as known strong emission or absorption lines (eg, H$\alpha$) or characteristic breaks (the Lyman break and the 4000 \AA\ break). If an object has more than one detected emission line, the wavelength ratio of the two can identify the spectral lines, whose observed wavelengths in turn pinpoint the redshift. However, not all objects will have detectable emission lines, and emission line signal-to-noise ratio will tend to decrease as the measured object gets fainter. Without emission lines, the location of a break (where the continuum flux level changes significantly) becomes the primary method of redshift identification \citep{ste96}.

This is most easily and accurately accomplished with the $R \sim 1000-2000$ spectra offered by ground-based instruments, but for increasingly distant and faint objects, ground-based spectroscopy becomes untenable. The break can also be detected in flux changes in photometric measurements. Photometric redshift fitting codes such as BPZ \citep{ben00}, Hyper-Z \citep{bol00}, LePhare \citep{ilb06}, and EAZY \citep{bra08} accomplish this by fitting broadband measurements against sets of template galaxy spectra. However, since the spectral coverage of a typical photometric band can cover $\sim$1000 \AA, this method lacks sufficient observations to fully constrain the fit, and thus is prone to significant systematic errors in the redshift identification.

By combining low- to mid-resolution HST grism spectra with ground-based broadband photometry, \citet{rya07} was able to achieve a fractional standard deviation of $\Delta z/(1+z_{spec}) \sim 0.04$, where $\Delta z$ is the difference between the grism calculated redshift and the ground-based spectroscopic redshift. This made a noticeable improvement over the purely photometric redshifts, which measured $\Delta z/(1 + z_{spec}) \sim 0.05$. This demonstrated that the addition of grism data could provide significant improvement in the calculations of redshifts for faint objects by identifying spectral features, and we find an improvement of the same order with this method. Similar methods were developed with G141 grism data in the 3D-HST survey \citep{bra12,mom16}.

In this paper, we present the catalog of SPZs developed for the Faint Infrared Grism Survey (FIGS) and our analysis of its quality. In \S2, we describe the observations and data reduction methods for the FIGS spectra. In \S3, we describe the computation of SPZs using EAZY. In \S4, we present our results, and measure the redshift accuracy in comparison to ground-based spectroscopic redshifts. In \S5, we explore the applications of this method to the study of LSS, and in \S6 we address the implications of this study for future surveys. We summarize our conclusions in section \S7.  Throughout this paper, we use AB magnitudes, and $\Lambda$-CDM cosmology with $H_0=70.0$ km s$^{-1}$Mpc$^{-1}$, $\Omega_m$=0.27, and $\Omega_\lambda=0.73$.

\section{Data and Methods}

\subsection{FIGS Observations and Spectral Extraction}

\subsubsection{Survey Description}

The Faint Infrared Grism Survey (FIGS, HST/Cycle 22, ID:13779, PI S. Malhotra) used the HST WFC3-G102 infrared grism to obtain deep slitless spectroscopy of $\sim$ 6000 galaxies. FIGS achieved 40-orbit depth in 4 fields, designated GN1, GN2, GS1, and GS2 (see Table 1 for coordinates of each field). Objects in each field were observed in 5 different 8-orbit position angles (PAs) in order to mitigate contamination of the spectra by overlapping spectra from nearby objects. Each PA covers a 2.05'x2.27' field of view. The area of coverage in each field from which we derive the spectra used for SPZs is given in Table 1, for a total area of 17.7 square arcminutes. 

\begin{table}
\begin{center}
\caption{A description of the four FIGS fields.}
\begin{tabular}{cccc}
\tableline
Field & RA & Dec & Area\footnotemark[1]  \\
\tableline
GN1 & 12:36:41.4670 & +62:17:26.27 & 4.51 \\
GN2 & 12:37:31.0234 & +62:18:26.91 & 5.06\\
GS1\footnotemark[2] & 03:32:40.9514 & --27:46:47.92 & 4.09 \\
GS2 & 03:33:06.4675 & --27:51:21.56 & 4.02 \\
\tableline
\end{tabular}
\scriptsize {
  \begin{tablenotes}
    \item[]$\rm ^a$ Measured in arcmin$^2$. \\
  \item[]$\rm ^b$ The HUDF. \\
  \end{tablenotes}}
\end{center}
\end{table}

\subsubsection{1D and 2D Spectral Extraction}

The FIGS G102 data reduction is described in detail in \citep{pir17}. FIGS data were reduced in a manner that loosely follows the method used for GRAPES and PEARS, previous HST grism surveys \citep{pir04}. First, we generated a master catalog of sources from deep CANDELS mosaics in the F850LP filter in ACS and the F125W and F160W filters in WFC3 (approximately the z, J, and H bands) \citep{gro11, koe11}. These mosaics provided absolute astrometric reference points for the FIGS F105W direct images and G102 dither exposures, along with individual FIGS F105W images. Contamination and background measurements were subtracted, and then 2D spectra were extracted using a process similar to that provided by the aXe extraction software \citep{kum09}. The final product is a set of multi-extension FITS files that each contain the spectrum of a science object, an error estimate, the object's contamination model, and its effective exposure map. Sources were extracted down to F105W $< 29.0$ mag.


One dimensional extractions were created from the 2D-extractions using two methods: Non-weighted extraction and Optimal extraction. The results in this paper were obtained using spectra made via Optimal extraction, which follows a non-iterative version of the algorithm described in \citep{hor86}. We used the simulated version of the 2D dispersed spectrum of the source to determine the expected profile of the spectrum as a function of wavelength. This profile was normalized to unity in the cross dispersion direction and used as the extraction weight. This extraction weight was then used in combination with the 2D contamination subtracted 2D data, to produce an optimally extracted 1D spectrum. The optimal extraction has the advantage of producing higher S/N spectra with improved flux calibration, but only when the extraction weights (derived from the imaging data) are accurate.


\subsubsection{PA Combination}

\begin{table*}
  \centering
  \caption{A description of the spectra samples of the four FIGS fields.}
\begin{tabular}{cccccccc}
\tableline
Field & Initial\footnotemark[1] & Not Matched\footnotemark[2] & Too Faint\footnotemark[3] & Low $\mathcal{N}$\footnotemark[4] & Bad Synthetic Flux\footnotemark[5] & Aperture Ratio\footnotemark[6] & Final\footnotemark[7] \\
\tableline
GN1 & 1913 & 263 & 812 & 241 & 21 & 25 & 551 \\
GN2 & 1003 & 84 & 161 & 222 & 20 & 50 & 466 \\
GS1 & 3106 & 390 & 1917 & 241 & 11 & 32 & 515 \\
GS2 & 2623 & 819\footnotemark[8] & 1223 & 194 & 13 & 27 & 347 \\
Total & 8645 & 1556 & 4113 & 898 & 65 & 134 & 1879 \\
\tableline
\end{tabular}
\scriptsize {
\begin{tablenotes}
\item[]$\rm ^a$ The number of combined 1D spectra in each field, before any quality cuts have been made. \\
\item[]$\rm ^b$ The number of objects with spectra without additional matched photometric data (See \S 2.2.1). \\
\item[]$\rm ^c$ The number of objects with spectra and matching photometry, but with F105 $> 26.5$ mag. \\
\item[]$\rm ^d$ The number of objects with spectra and matching photometry, but with a net spectral significance less than 10 (See \S 2.2.2). \\
\item[]$\rm ^e$ The number of objects meeting previous criteria for which we were unable to calculate a synthetic F105W magnitude. \\
\item[]$\rm ^f$ The number of objects meeting previous criteria but rejected for having a large aperture correction or less than 90\% full coverage. (See \S 2.2.3) \\
\item[]$\rm ^g$ The number of objects in each field that pass all quality criteria and are used in the final SPZ sample.
\item[]$\rm ^h$ GS2 lacks deep WFC3 imaging at some roll angles. See \S 2.2.1.
  
\end{tablenotes}}
\end{table*}

Because the total data for each of the four fields comes in 8-orbit segments for separate PAs, one must consider how to merge the data in a set of contamination-subtracted PAs in order to achieve the best signal for the largest number of objects. The observed spectra are the convolution of the light profile of the object with its spectrum, and large differences in this light profile between different PAs (for example, in the cases of elliptical or irregular galaxies) will result in spectra that disagree strongly near the edge of the bandpass of the grism. They will also have continuum fluxes that are in disagreement, as the spectrum is smoothed by different amounts. We derived an object-specific spectral response for each source by dividing the extracted 1D data by the extracted 1D simulated data, and by the spectral energy distribution used to generate the FIGS simulations, which were generated from the available FIGS broad band photometry. The result is a normalized spectrum, which can be scaled back to the observed F105W photometry. These steps insure that the 1D spectra of extended sources are accurately flux-calibrated and avoid the issue of having a point-source sensitivity function applied to an extended object.

The FIGS spectra were flux calibrated using object specific sensitivity functions and then combined. For each wavelength bin, the inverse variance of the singlePA spectra were used as weights to compute the weighted mean and standard deviation of the weighted mean. An iterative 3-sigma rejection was used to remove outlier single PA spectral bins.





\subsection{The Sample}

The sample of FIGS spectra used are described in detail in the following section, the results of which are summarized in Table 2.

\begin{figure}
  \includegraphics[width=0.5\textwidth]{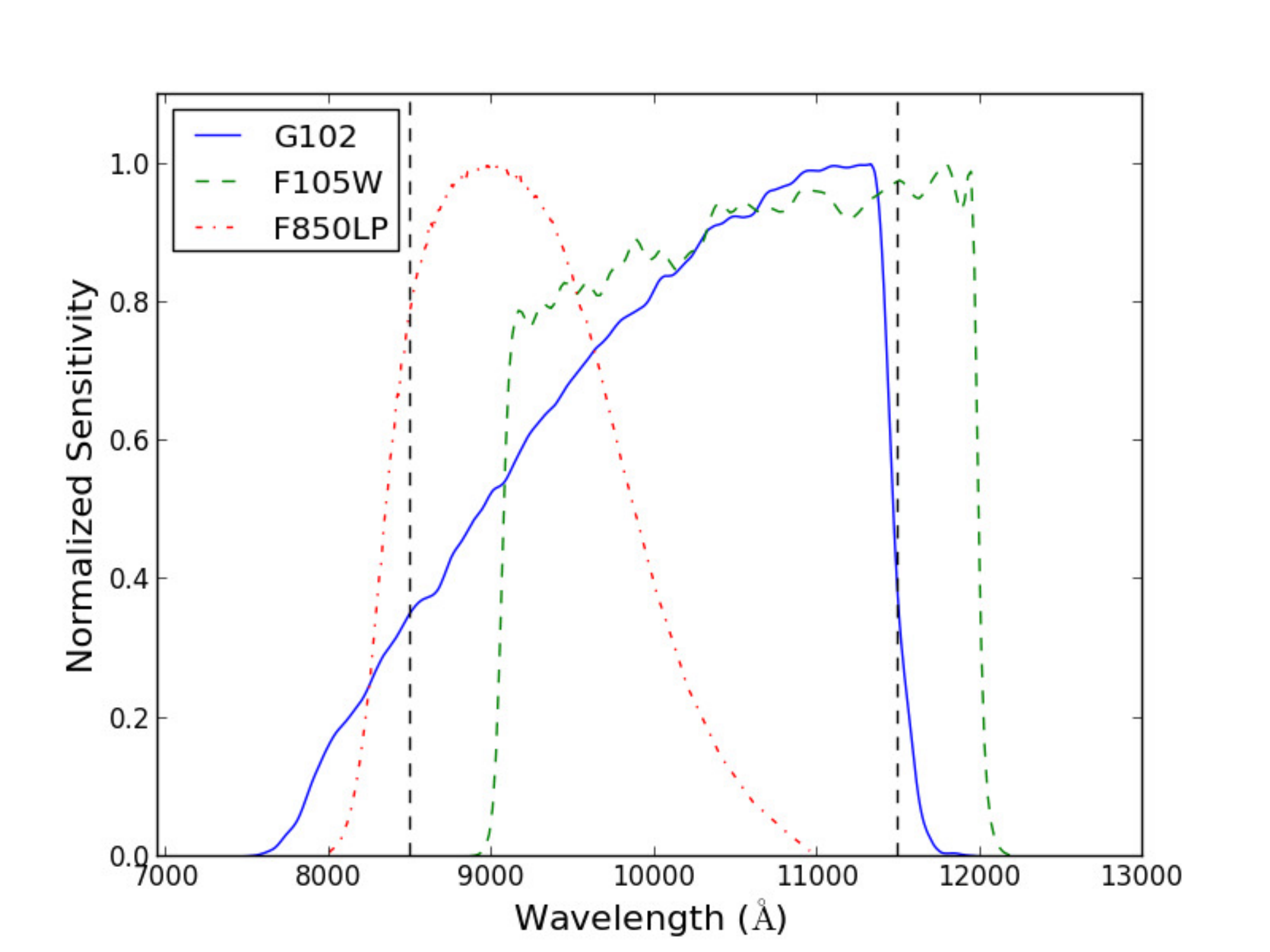}
  \caption{The sensitivity curves for the WFC3/G102 grism, as seen in \citet{kun11}, and the F105W and F850LP filters. The dashed vertical lines show the cutoffs for grism data used in the construction of redshifts. The curves have been normalized to their maximum sensitivity.}
\end{figure}

\subsubsection{Broadband Photometric Data}

Given that the wavelength coverage of FIGS spectra is limited to the 8500-11500 \AA\ wavelength range, additional data are often useful for constraining the parameters of the redshift fit. To extend the spectral range of the fit, we supplement the FIGS spectra and photometry with optical and mid-IR broadband photometry available from previous surveys: GOODS \citep{dic03, gia04}, CANDELS \citep{koe11}, MODS \citep{kaj11}, SEDS \citep{ash13}, and HHDFN \citep{ste03,cap04}. See Tables 3 and 4 for specific details on the broadband data used. We obtained these measurements from the combined, PSF-matched catalogs produced by 3D-HST \citep{ske14}.

\begin{table*}
\begin{center}
\caption{Sources of Broadband Photometry}
\begin{tabular}{cccc}
\tableline
Filters & Telescope/Instrument & Survey & Reference \\
\tableline
U$^1$ & KPNO 4m/Mosaic & Hawaii HDFN & \citep{cap04} \\
U, R$^2$ & VLT/VIMOS & ESO/GOODS & \citep{non09} \\
U38, B, V, R$_c$, I$^2$ & WFI 2.2m & GaBoDs & \citep{hil06, erb05} \\
G, R$_s^1$ & Keck/LRIS & Hawaii HDFN & \citep{ste03} \\
F435W, F606W, F775W, F850LP & HST/ACS & GOODS & \citep{gia04} \\
B, V, I$_c$, z'$^1$ & Subaru/Suprime-Cam & Hawaii HDFN & \citep{cap04} \\
F140W & HST/WFC3 & 3D-HST & \citep{bra12} \\
F125W, F160W & HST/WFC3 & CANDELS & \citep{koe11} \\
J, H, K$_s^1$ & Subaru/MOIRCS & MODS & \citep{kaj11} \\
J, H, K$_s^2$ & VLT/ISAAC & ESO/GOODS, FIREWORKS & \citep{ret10, wuy08} \\
3.6, 4.5 $\mu$m & Spitzer/IRAC & SEDS & \citep{ash13} \\
5.8, 8 $\mu$m & Spitzer/IRAC & GOODS & \citep{dic03} \\
\tableline
\end{tabular}
\begin{tablenotes}
  \item[]$\rm ^1$ Northern fields
  \item[]$\rm ^2$ Southern fields
\end{tablenotes}
\end{center}
\end{table*}

\begin{table}
\caption{Broadband Photometry Depths}
\begin{tabular}{ccccc}
\tableline
Filter & $5\sigma$ Depth (N) & $5\sigma Depth$ (S) & $\lambda_{central}$ (\AA) & Width(\AA) \\
\tableline
U & 26.4 & - & 3593 & 721 \\
R & 26.2 & 27.5 & 6276 & 1379 \\
U38 & - & 25.7 & 3637 & 475 \\
B & 26.7 & 26.9 & 4448 & 1035 \\
V & 27.0 & 26.6  & 5470 & 993 \\
R$_c$ & - & 26.6 & 6517 & 1600 \\
I & 25.8 & 24.7 & 7671 & 1489 \\
G & 26.3 & - & 4751 & 940 \\
R$_s$ & 25.6 & - & 6819 & 1461 \\
F435W & 27.1 & 27.3 & 4318 & 993 \\
F606W & 27.4 & 27.4 & 5919 & 2225 \\
F775W & 26.9 & 26.9 & 7693 & 1491 \\
F850LP & 26.7 & 26.5 & 9036 & 2092 \\
z' & 25.5 & - & 9028 & 1411 \\
F140W & 25.9 & 25.6 & 13924 & 3760 \\
F125W & 26.7 & 26.1 & 12471 & 2876 \\
F160W & 26.1 & 26.4 & 15396 & 2744 \\
J & 25.0 & 25.1 & 12517 & 1571 \\
H & 24.3 & 24.5 & 16347 & 2686 \\
K$_s$ & 24.7 & 24.4 & 21577 & 3044 \\
IRAC1 & 24.5 & 24.8 & 35569 & 7139 \\
IRAC2 & 24.6 & 24.8 & 45020 & 9706 \\
IRAC3 & 22.8 & 23.0 & 57450 & 13591 \\
IRAC4 & 22.7 & 23.0 & 79158 & 27839 \\
\tableline
\end{tabular}
\end{table}

Because FIGS is targeting faint objects, there are some detected FIGS spectra for which we are unable to find matching photometry. These objects amount to $\sim 10$\% of the objects in GN1, GN2, and GS1 (HUDF), and a number of these spectra would later be rejected for other reasons (eg, low spectral signal). To cut down on spectra for which we could not expect a useful signal, we then applied a magnitude cutoff at F105W $< 26.5$ mag.

About a third of the GS2 spectra, however, lack the matched photometry described in Table 3. GS2 is centered on one of the HUDF parallel fields outside GOODS-S, and parts of the FIGS WFC3-IR field lie outside the deepest WFC3 imaging data for some roll angles. If we check other criteria to identify usable spectra before photometric matching first, we find that there is an average of 40 objects per field which have viable FIGS spectra but do not have an existing match in the catalogs compiled by 3D-HST. Redshifts can still be computed successfully for these objects, though they lack significant constraints in the NIR and IRAC bands, potentially increasing errors, so we exclude such objects from the calculations in this paper. See Section 3 for a further discussion.

\subsubsection{Net Significance}

As described in the instrument calibration report \citep{kun11}, the WFC3 G102 grism achieves a maximum throughput of $41\%$ at 11000 \AA\ and provides $\geq 10\% $ system throughput in the wavelength range $8035-11538$ \AA\ (see Figure 1). At longer wavelengths, the throughput declines rapidly, making this the effective useful range of the spectra. We assess the content of the individual spectra by computing the net significance \citep{pir04}, which is determined by reordering the resolution elements in order of descending signal-to-noise ($S/N$) ratio, and then iteratively computing a cumulative $S/N$ ratio from the current element and all lower elements. This continues until a maximum $S/N$ ratio is computed. After the cumulative $S/N$ ratio turns over, adding additional data will not increase the $S/N$ ratio. The maximum is then the net significance of the spectrum, $\mathcal{N}$. As described in \citet{pir17}, a simulated FIGS spectrum with a continuum level of $3\sigma$ per bin is expected to have $\mathcal{N} \approx 4.5$. It is possible to obtain an artificially high value of $\mathcal{N}$ for an object if there is unaccounted contamination or errors in the level of background subtration. In order to avoid such objects and to ensure we used only objects with high signal, we imposed a netsig cutoff of $\mathcal{N} > 10$. This eliminates 7-20\% of the initial number of objects in each field. Though it is still possible for contaminated or otherwise low-signal objects to bypass this cutoff, such objects are likely to be caught by further quality cuts.






\subsubsection{Aperture Correction}

\begin{figure*}
\begin{tabular}{cc}
\includegraphics[width=0.5\textwidth]{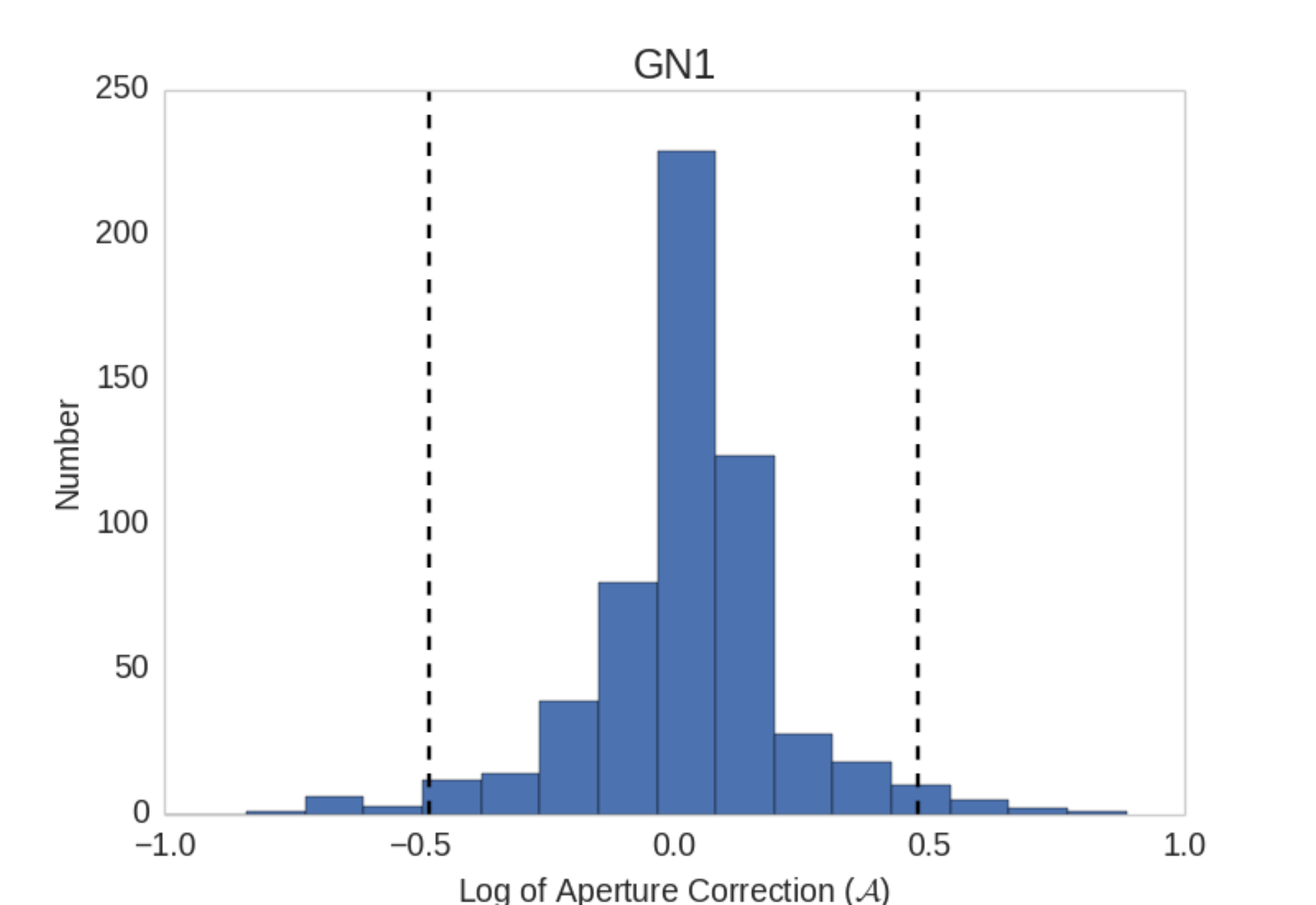} & \includegraphics[width=0.5\textwidth]{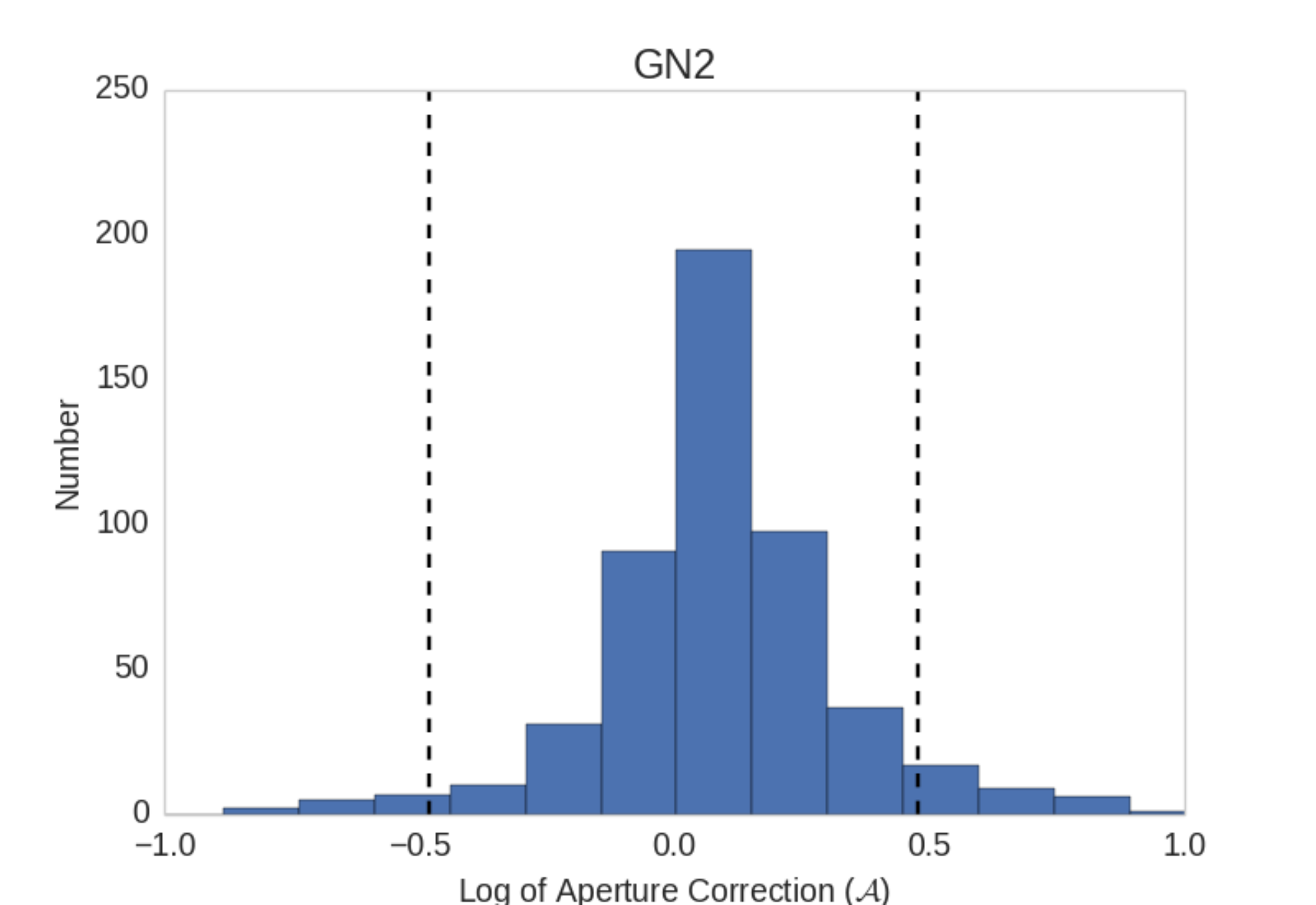} \\
(a) 1029 objects & (b) 742 objects \\[6pt]
\includegraphics[width=0.5\textwidth]{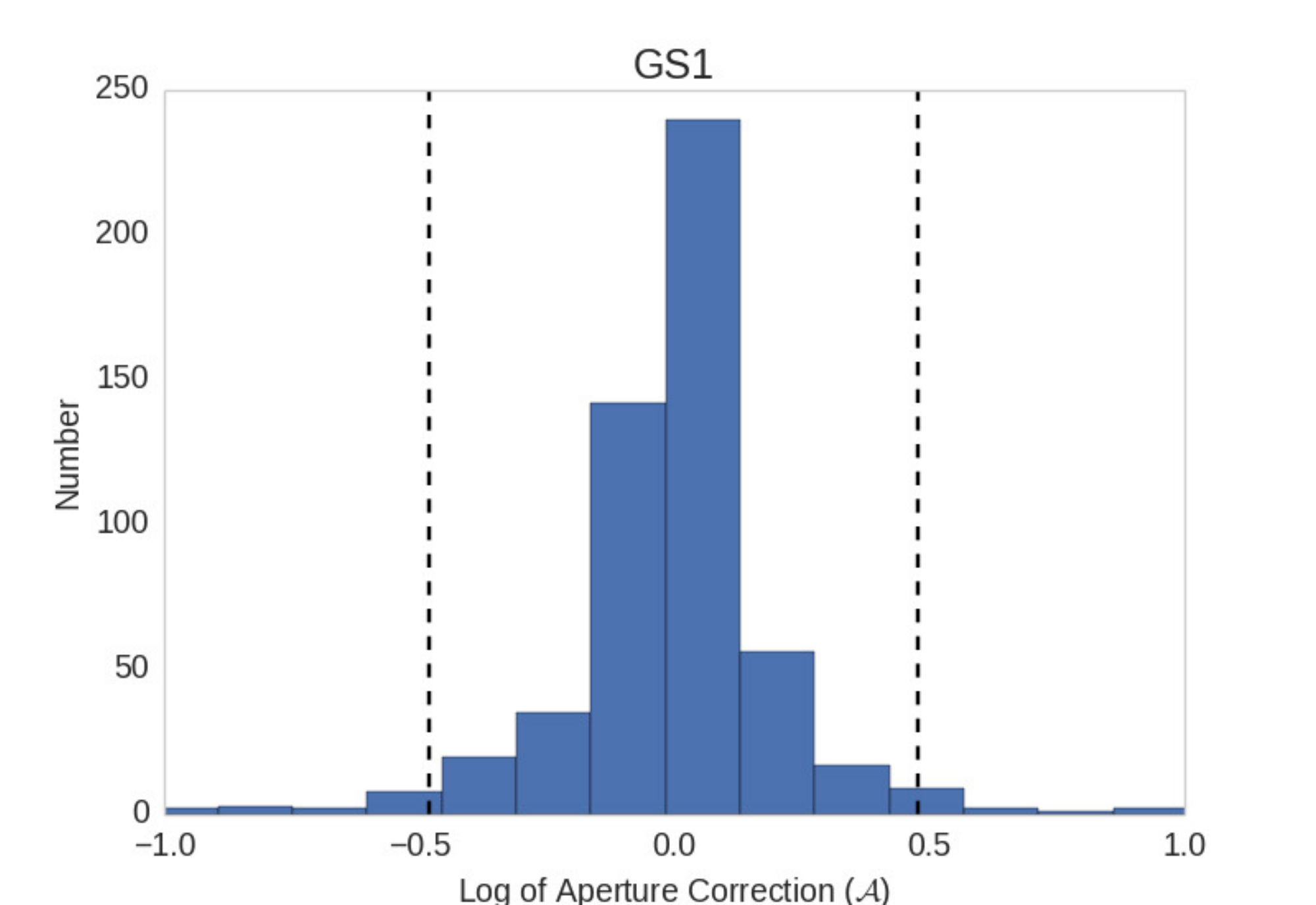} & \includegraphics[width=0.5\textwidth]{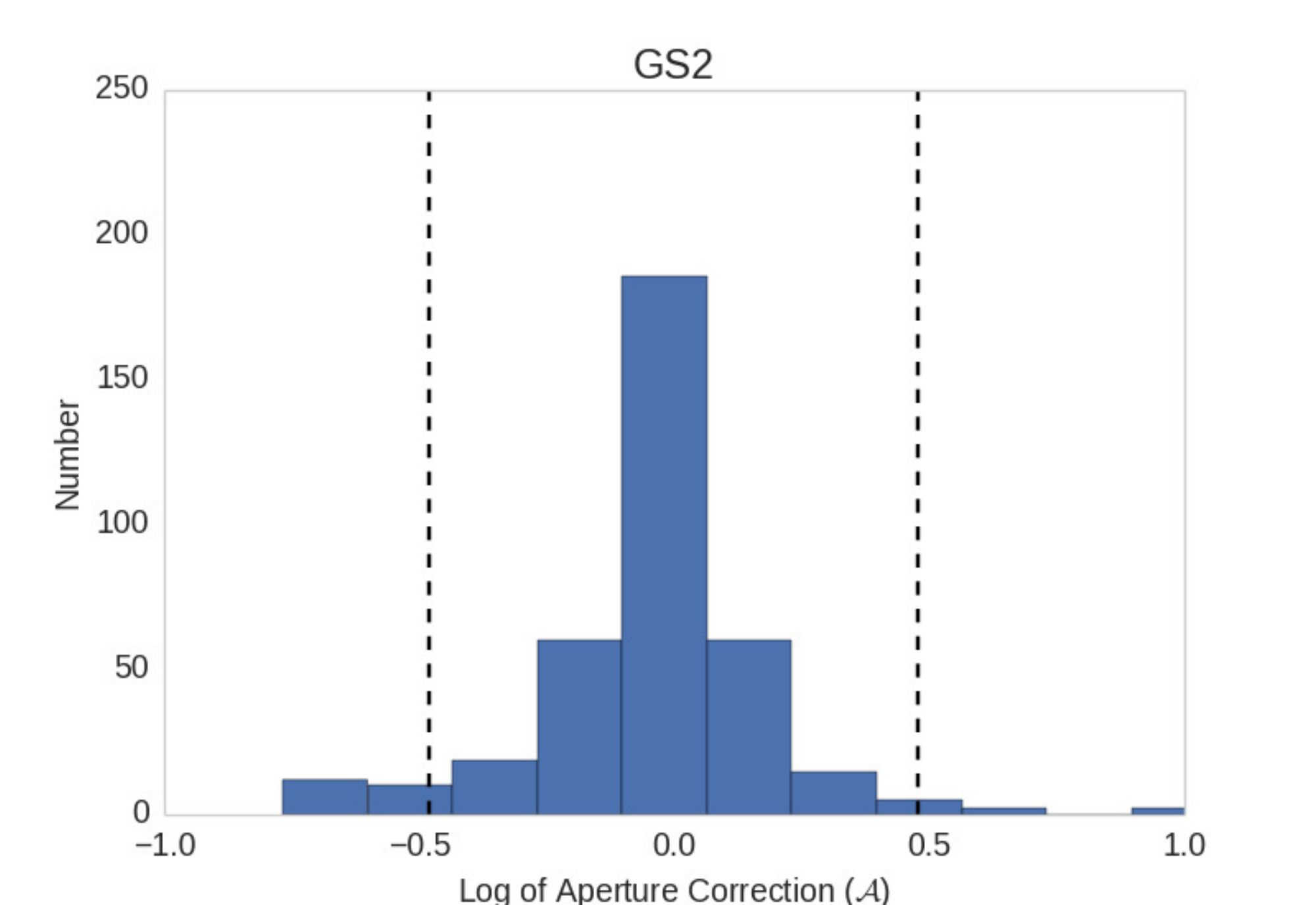} \\
(c) 1163 objects & (d) 805 objects \\[6pt]
\end{tabular}
\caption{The distributions of $\mathcal{A} = \log_{10} \left( \text{F105W(obs)} / \text{F105W(synth)} \right)$ in each field: (a) GN1 (b) GN2 (c) GS1 (d) GS2. These histograms are given in terms of the log of the ratio of the measured broadband F105W flux to the synthetic F105W flux, so a value of 0 indicates a 1:1 ratio. Bin widths are consistent for all four plots. To get a sense of the shape of each distribution, a Gaussian was fit to the distributions, and the FWHM calculated.}
\end{figure*}

\begin{figure*}
\begin{tabular}{cc}
\includegraphics[width=0.5\textwidth]{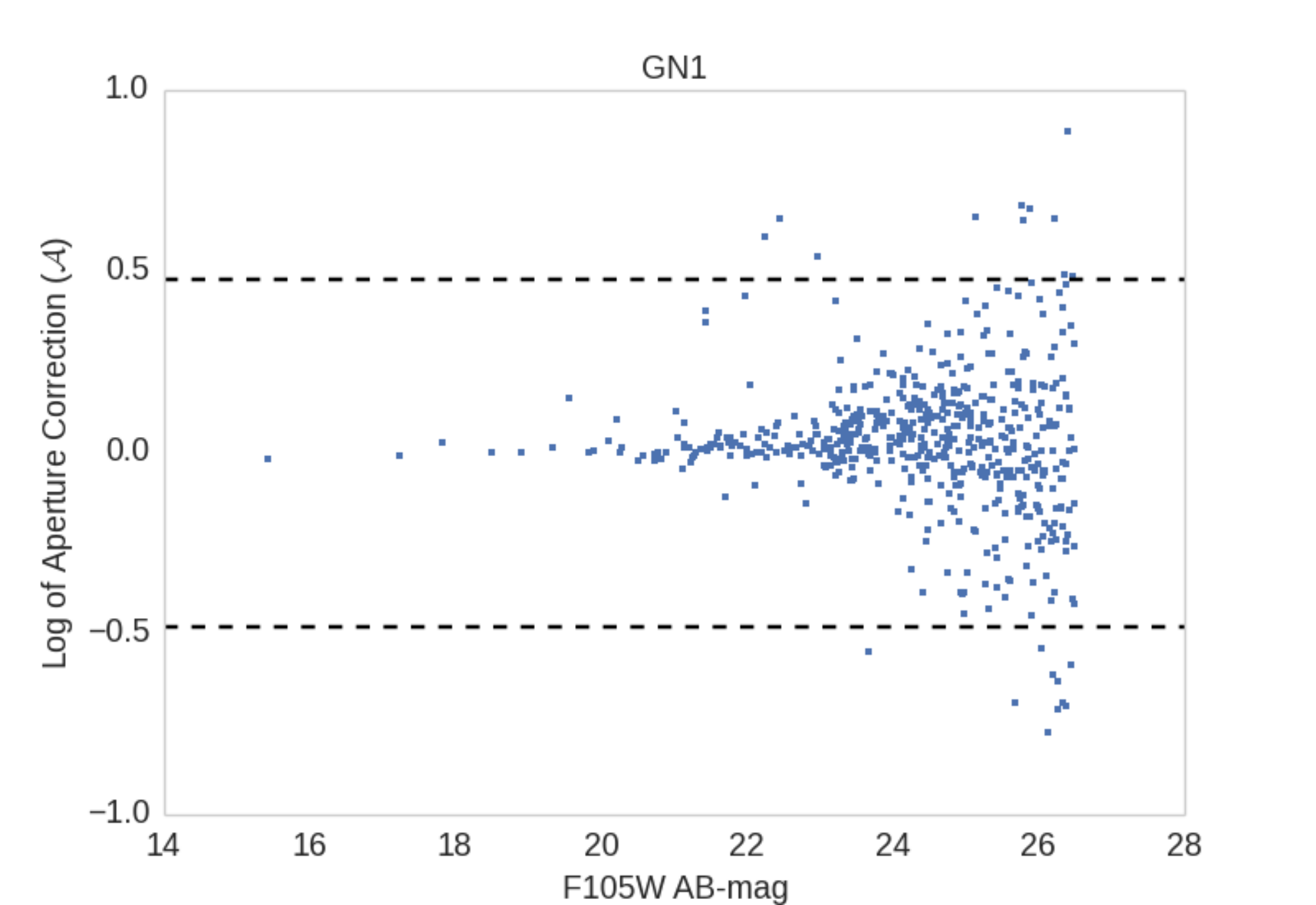} & \includegraphics[width=0.5\textwidth]{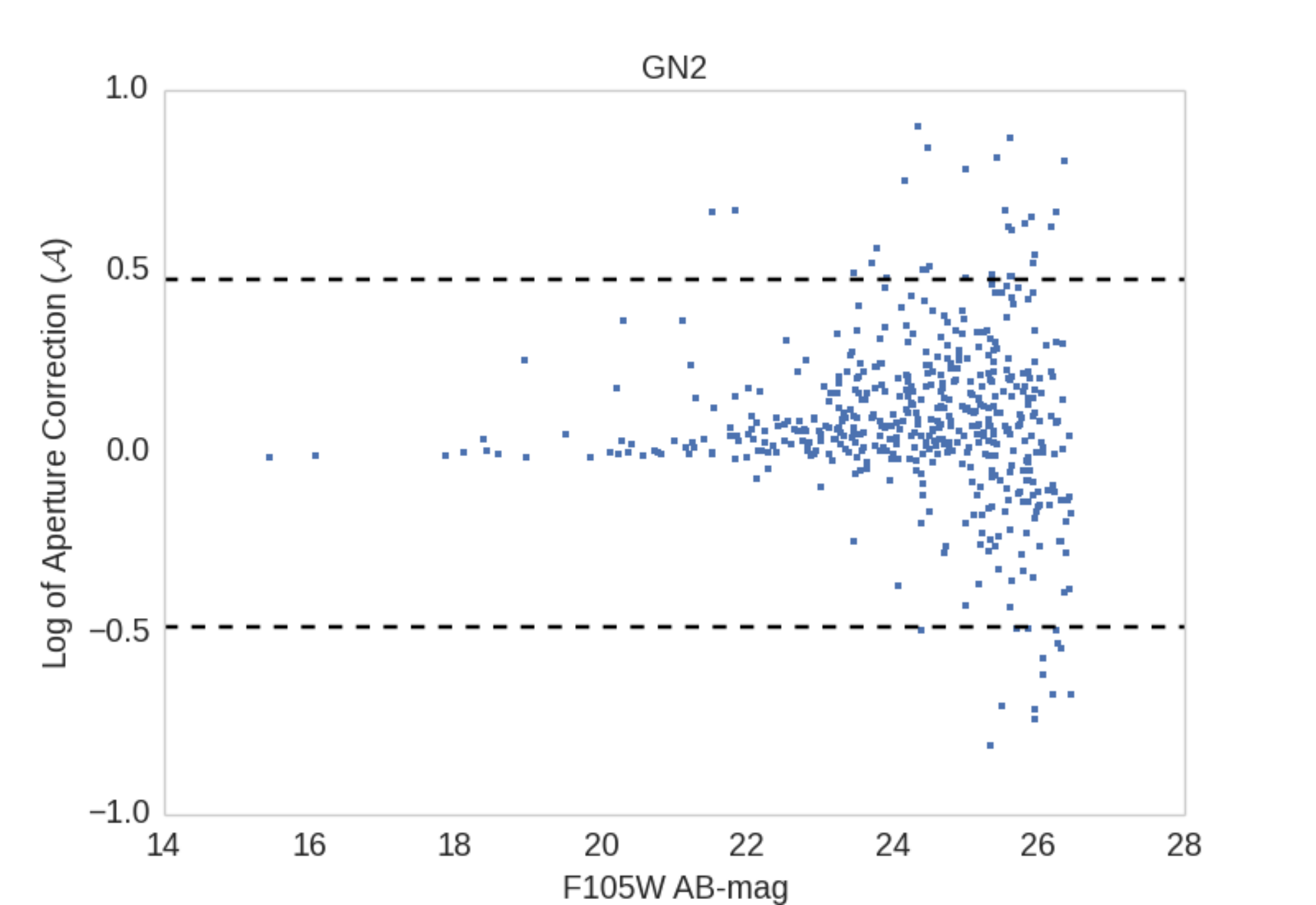} \\
\includegraphics[width=0.5\textwidth]{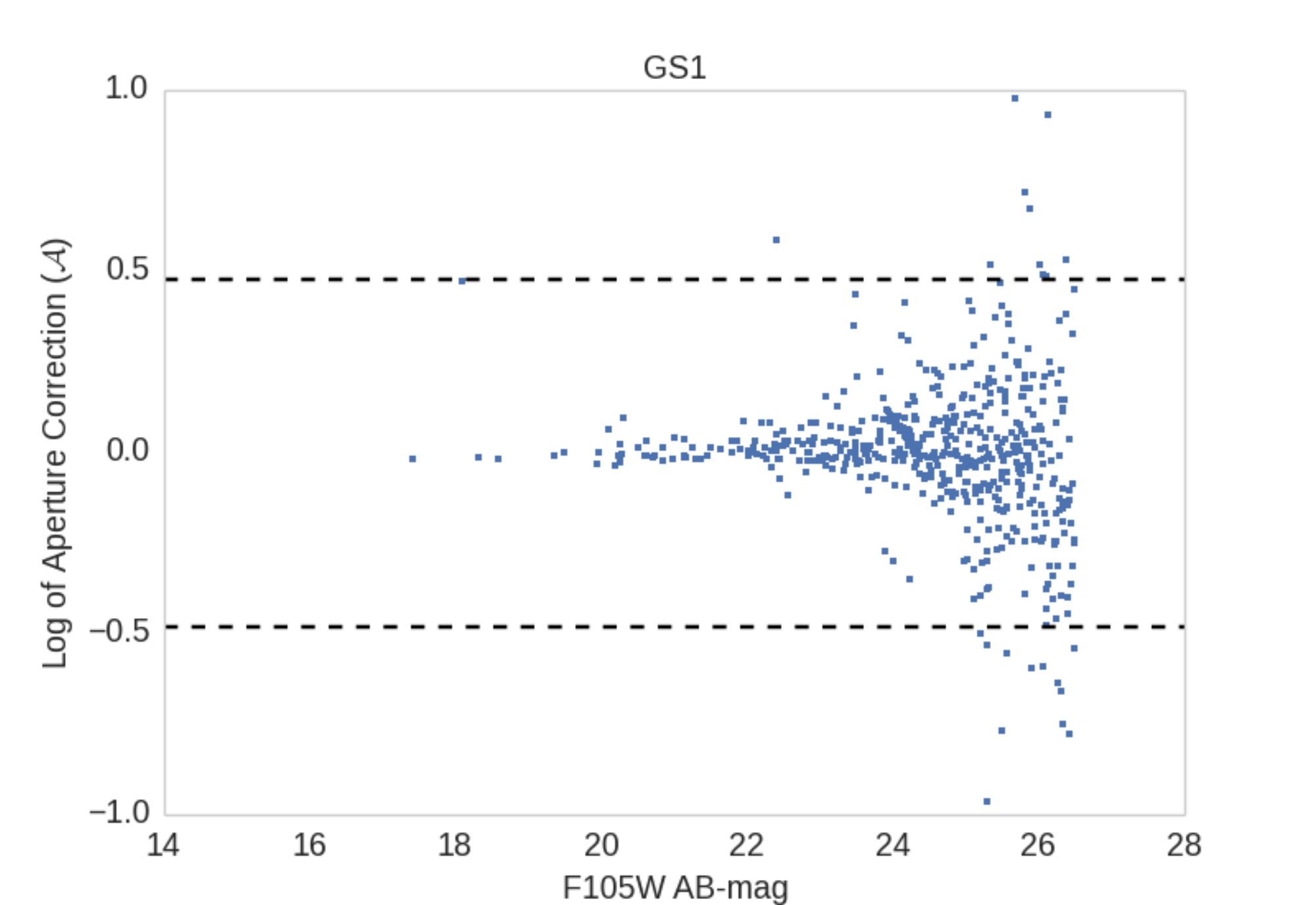} & \includegraphics[width=0.5\textwidth]{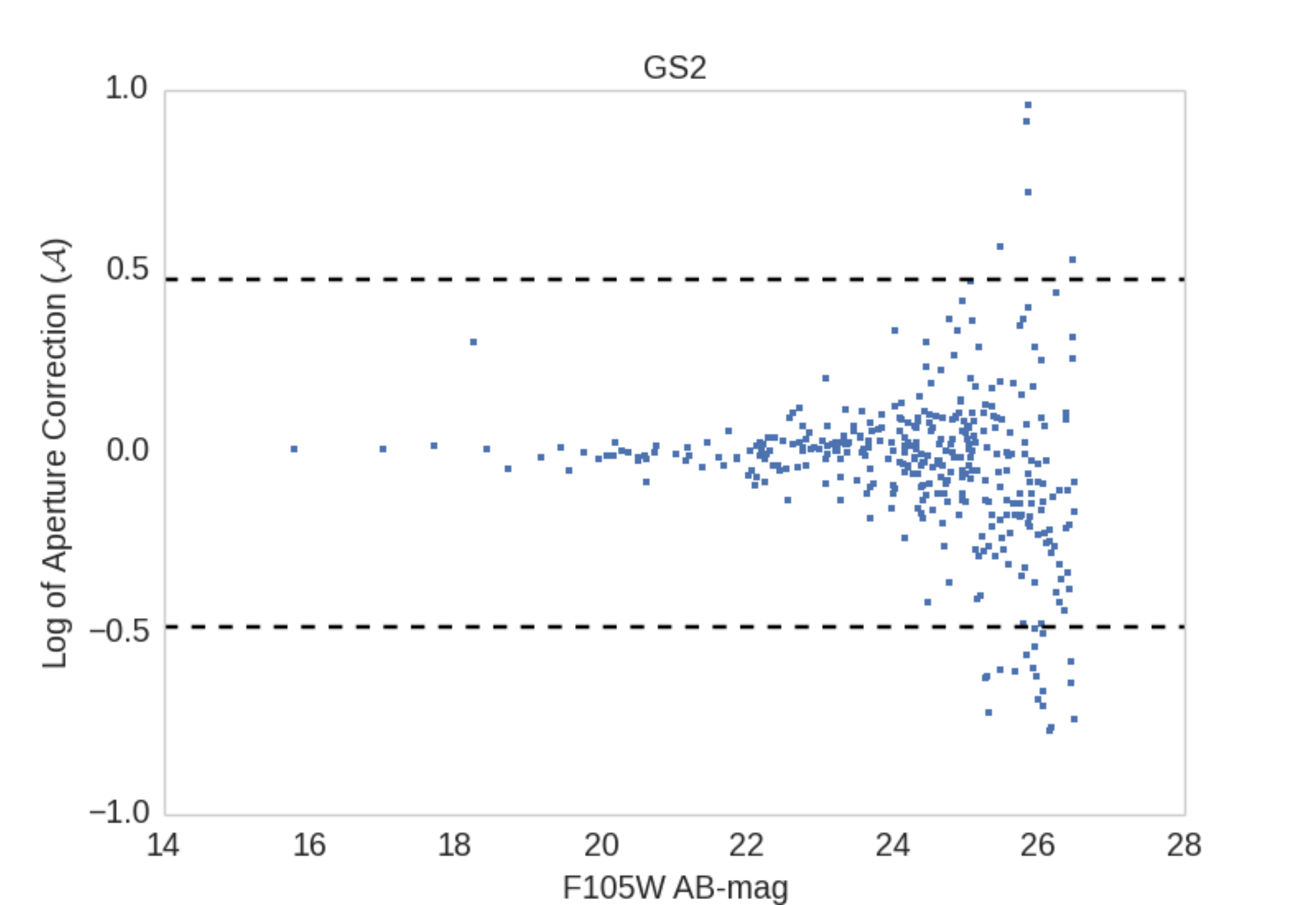} 
\end{tabular}
\caption{The aperture correction distributions ($\mathcal{A} = \log_{10} \left( \text{F105W(obs)} / \text{F105W(synth)} \right)$) as a function of F105W magnitude in each field: (a) GN1 (b) GN2 (c) GS1 (d) GS2. The aperture correction is given in terms of the log of the ratio of the measured broadband F105W flux to the synthetic F105W flux, so a value of 0 indicates a 1:1 ratio.}
\end{figure*}

Because the spectra will need to be combined with photometry in order to cover the wavelength range needed for a redshift fit, the flux values in the spectra need to be scaled to match those of photometric images. To do this, we define an aperture correction, which is the flux ratio between a photometric flux and a synthetic flux calculated from the spectrum in the same band, as described in \citet{rya07}. Figure 1 shows the grism throughput curve plotted against the two nearest HST broadband filters: F850LP and F105W. Both filters extend past the usable wavelength range of the grism, but the F105W band has the closest filter profile correspondence to the grism throughput, since the F850LP sensitivity anticorrelates with that of the grism, resulting in a much broader distribution of aperture corrections. Consequently, we defined the aperture correction $\mathcal{A}$ in terms of a synthetic F105W band, calculated by integrating over the product of grism spectra with the F105W filter curve:

\begin{equation}
  \mathcal{A} = \log_{10} \left( \frac{\text{F105W(obs)}}{\text{F105W(synth)}} \right)
\end{equation}
  
\noindent If for some reason a synthetic F105W flux cannot be calculated from the spectrum (usually if oversubtraction of contamination left most of the fluxes negative), that spectrum is rejected for SPZ use.

The distribution of the aperture corrections in each field is displayed in Figure 2. The widths of the aperture correction distributions are a function of the spectral extraction method, the broadband apertures, and spectral contamination from nearby objects. However, as noted above, the F105W filter profile goes significantly redder than the G102 wavelength coverage. If an object spectrum in that region is not flat, then the aperture correction produced is likely to be quite large.

In all the fields, the distribution peaks sharply near 0, and the distributions feature a negative tail, indicating a tendency of the synthetic F105W measurements to exceed that of the HST photometry. The shape of this distribution becomes clearer when looking at Figure 3, which displays the aperture correction as a function of F105W magnitude (a proxy for the brightness of the spectrum). The aperture corrections only begin to strongly diverge from 0 for fluxes fainter than $\sim 25.5$ mag, with the most deviant objects typically found at the very faintest magnitudes. This is likely a function of contamination, which will make up a larger fraction of the total measured flux in an object with a faint true brightness. Consequently, accurately estimating the contamination in such objects is more difficult, increasing the likelihood that a faint object will retain some contaminating flux. Faint contaminated objects are therefore more likely to have synthetic F105W measurements that are significantly larger than the broadband measurement, hence the negative tail in the distribution. To avoid the influence of such objects, we imposed a cutoff in aperture correction where any object with a ratio off from unity by a factor of 10 or more is rejected. Across all four fields, an average of 6\% of the initial sample was rejected for this reason. This correction does not account for any objects whose continuum slope has been altered by the presence of unaccounted background or contamination, which may pass through the aperture correction if the overall integrated flux doesn't change much. Such objects were later weeded out via visual inspection.

Some objects do not have a measurement in all 5 PAs, and some PAs don't measure the flux across the total G102 wavelength range. This usually does not cause any issues with the combination of the different PAs, but it may if the only PAs with data do not have overlapping coverage. This can result in combined spectra with gaps in the flux, and such spectra tend to produce very confused results in the redshift fit. These and any other spectra with missed contamination were rejected by visual inspection. These final removals typically amounted to $\sim$1\% of the initial sample.

\section{Redshift Estimation}

\subsection{Photometric Fitting Code}

To estimate the redshifts, we used EAZY \citep{bra08}, a public photometric redshift code. A systematic comparison of 9 different photometric redshift codes (including EAZY) across 11 different photometric redshift catalogs \citep{dah13} found that no particular code obtained significantly more accurate photometric redshifts compared to the others. Given this, we use EAZY for its modifiability that allows the simple inclusion of grism data alongside photometry.

Given a set of photometric points and corresponding errors, EAZY can iterate over a grid of redshifted spectral templates, calculating synthetic fluxes to compare to the measured photometry. The differences between the synthetic templates fluxes and the observed fluxes are used to define a $\chi^2$ statistic. This $\chi^2$ is minimized across the range of template spectra and redshifts to find the best-fit template and redshift. EAZY's template spectra are derived from a library of $\sim3000$ P\'EGASE spectra \citep{fio97}, including spectra with variable-strength emission lines. Using the method of non-negative matrix factorization \citep{bla07}, \citet{bra08} were able to reduce the large template set to a set of 5 basis templates that, in linear combination, can represent the full range of colors of the initial set. These basis templates, along with a dusty starburst template (a Calzetti extinction law is applied), make up the template set EAZY uses for fitting. 

Despite the breadth of this template set, there remains uncertainty in the spectral properties (eg, variations in dust extinction) that go into constructing template galaxies, which may result in mismatches between the templates and an observed galaxy. To account for this, EAZY provides a template error function to account for this uncertainty when fitting observations to the templates. The template error function provides a per-wavelength error in the template flux derived from the residuals of a large set of redshift fits. This allows the template set to accomodate observational spectral variations that aren't accounted for in the physics from which the templates are derived. 

Furthermore, to avoid degeneracies in the redshift fitting wherein the redshift probability distribution produces more than one peak, EAZY also provides a grid of magnitude priors in R- and K-bands, which assigns probabilities to measuring a band at a certain brightness at a given redshift, a technique first applied in \citet{ben00}. This typically reduces the incidence of catastrophic failures, where the difference between the grism redshift and spectroscopic redshift is more than 10\% of (1+$z_{spec}$), by providing a mechanism for avoiding wrong-peak selection in the case of a bimodal probability distribution. In our sample, we use the R-band magnitude prior, calculated from the observer-frame R-band flux.

\subsection{Inclusion of Grism Data}


\begin{figure*}
\includegraphics[width=\textwidth]{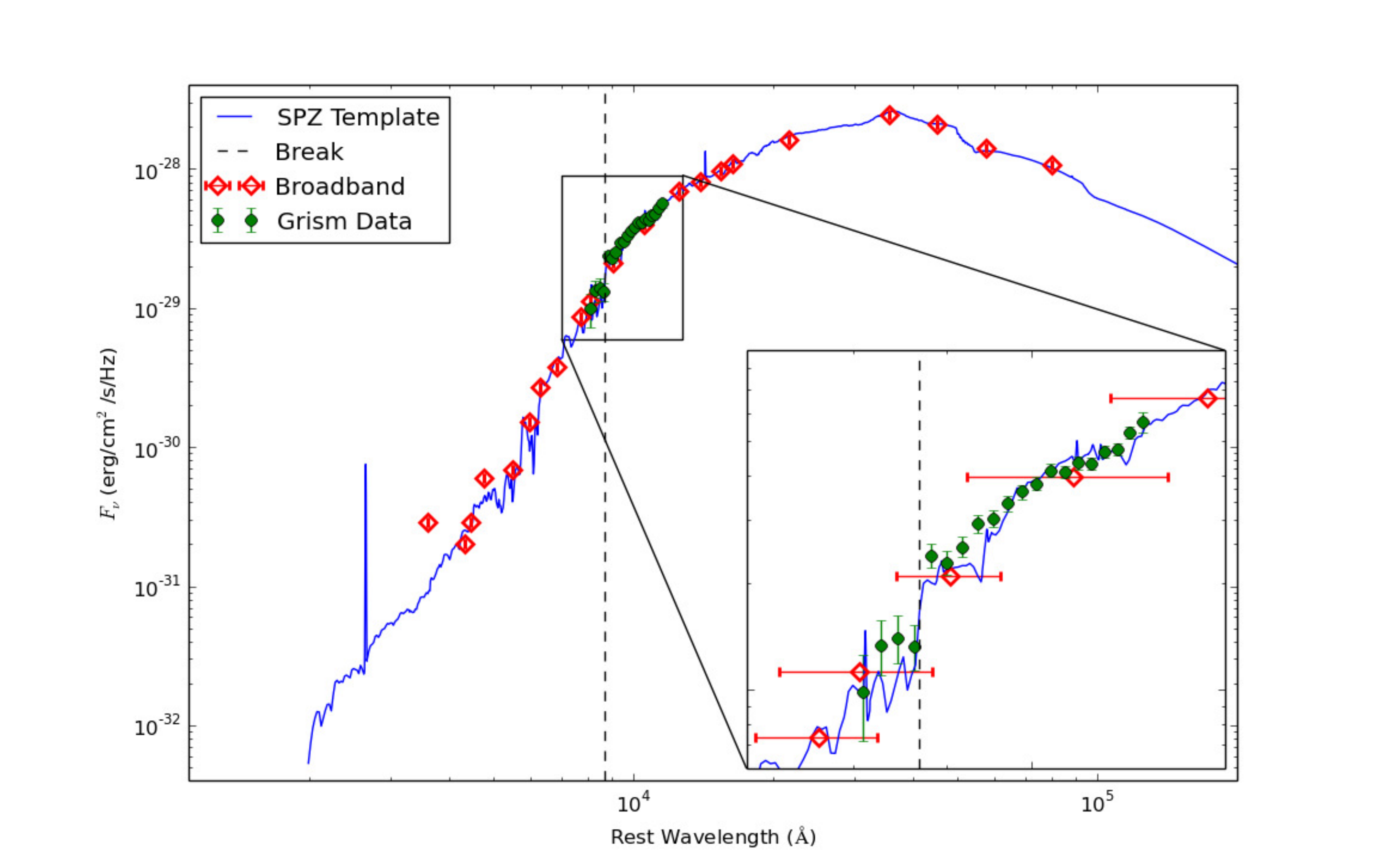}
\caption{Full: a spectral energy distribution (SED) of an example FIGS galaxy showing the EAZY input and output. Red points are FIGS and 3D-HST broadband data, and green points are the grism spectral data from FIGS. Both sets of points go into EAZY's calculations. In addition to the redshift, EAZY also outputs the template spectrum for which the $\chi^2$ is minimized (shown in blue). Inset: a close-up of the SED around the grism wavelength coverage. Vertical error bars represent the flux error (which is typically too small to see in the broadband), and horizontal error bars represent the effective width of the broadband filters. The wavelength width of the grism points is $\sim 140$ \AA\ . The dashed line marks the location of the 4000 \AA\ break at the predicted redshift. One can see the break precisely in the narrowband SED as well as in the output template. The input includes photometric points beyond the wavelength range depicted. The wavelengths of the plot were restricted in order to focus on the grism region.}
\end{figure*}

To include the FIGS spectra in a photometric fitting code, we followed the procedure in \citet{rya07} and reformatted the spectra into a series of narrow photometric bands that could be supplied to EAZY alongside broadband photometry. After passing the net significance and aperture correction procedures without rejection, an individual spectrum is divided into sub-samples along its operating wavelength range. The number (and therefore width) of these sub-samples can influence the results of the redshift fit. We experimented with a number of bins ranging from a few (width $\sim750$ \AA) to treating each grism element as its own bin (a width of 24.5 \AA), in order to obtain the best results. Typically, grism points derived from fewer, broader wavelength bins are better at avoiding errors introducted by problems in the combined spectra (eg, over-subtraction of contamination), since each bin will include a larger number of pixels, reducing the influence of one or two bad pixels. However, the redshift quality of the whole sample is best with a larger number of narrower bins, as this allows for the more precise location of breaks and emission lines. The cases where over-subtraction or other errors produce inaccurate results are few enough in number that they can be flagged individually, so we attempted the redshift fits with the narrowest grism bands. Bands narrower than a few grism pixels caused the fitting routine to stall and fail to produce a redshift fit. Consequently, we chose to proceed with grism bands $\sim$140 \AA\ wide (which corresponds to $\leq$ 22 spectral ``pixels'' per spectrum), which was the narrowest range to fit successfully. This may not result in any loss of improvement, as the typical spatial scale of objects in FIGS is a few pixels, so the spectra are smoothed to this extent anyway.

The flux in each of these subsamples is integrated to produce a new ``narrowband'' flux in each sample. These narrowband fluxes are written into an EAZY input catalog alongside FIGS and 3D-HST broadband photometry. EAZY is also given a ``filter profile'' for each narrowband in the form of a tophat function bound by the wavelength range of the grism band. EAZY provides the option to smooth the filter profiles $R(\lambda)$ by applying a Gaussian such that $\mathcal{R}_i = (1/b_i) \Sigma_j R(\lambda) \cdot G(\lambda_i, \lambda_j, \sigma)$ where $G$ is a Gaussian function and $b_i = \Sigma_j G$. After testing several cases, we obtain the best results with smoothing enabled with $\sigma=100$ \AA. We run EAZY on a redshift grid of $z=0.01-6.0$, which is the redshift range tested in EAZY's design, with $\Delta z = 0.01\cdot(1 + z_{prev})$.

Figure 4 shows an example of EAZY input and output for one of the FIGS objects exhibiting a particularly noticeable 4000 \AA\ break. The location of the break is more obvious at the higher resolution of the grism data, which confine it to a $\sim 100$ \AA\ wavelength range, as opposed to the $\sim 1000$ \AA\ coverage provided by the broadband photometry alone.

\section{Results}


\begin{figure*}
\begin{tabular}{cc}    
\includegraphics[width=0.5\textwidth]{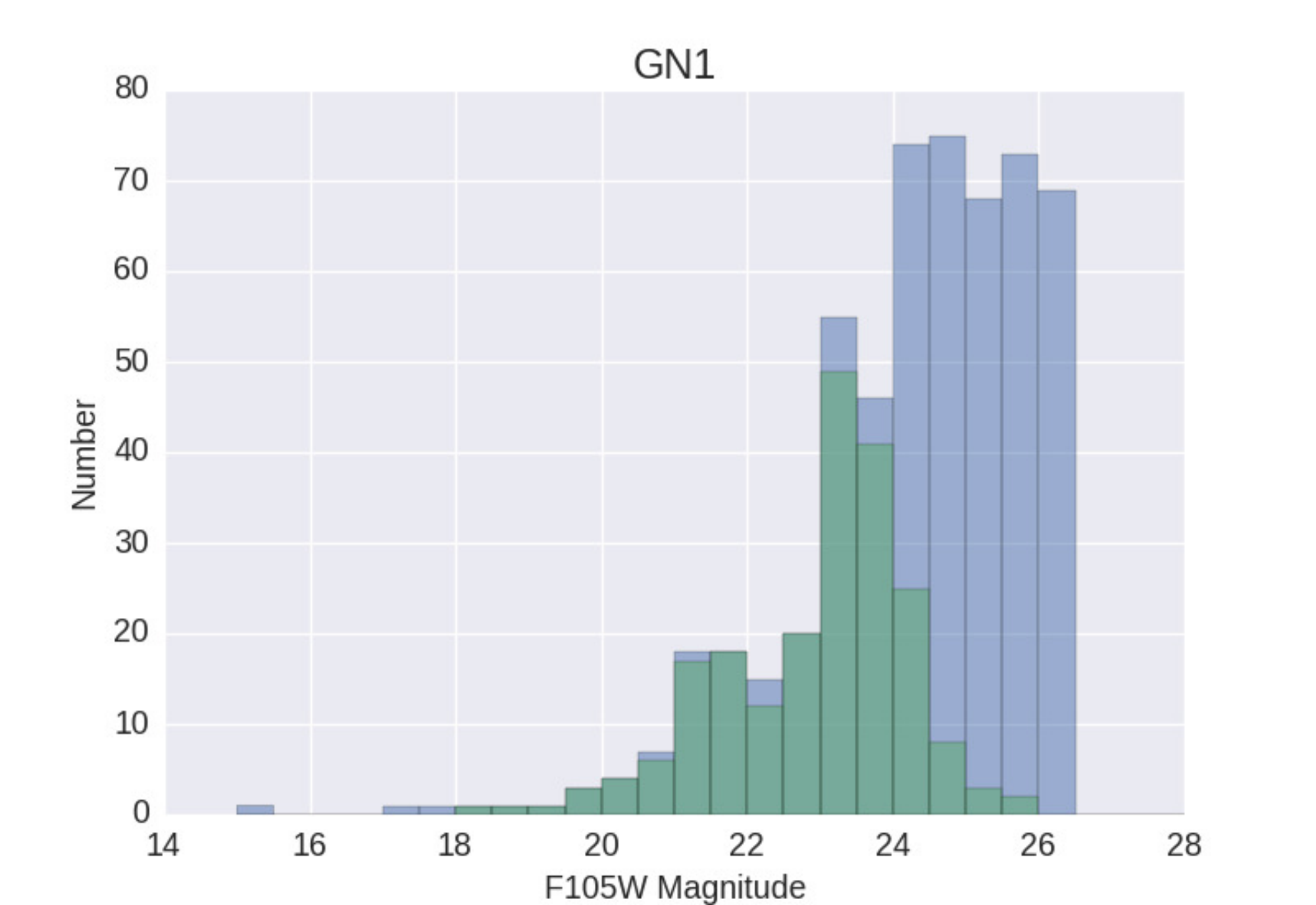} & \includegraphics[width=0.5\textwidth]{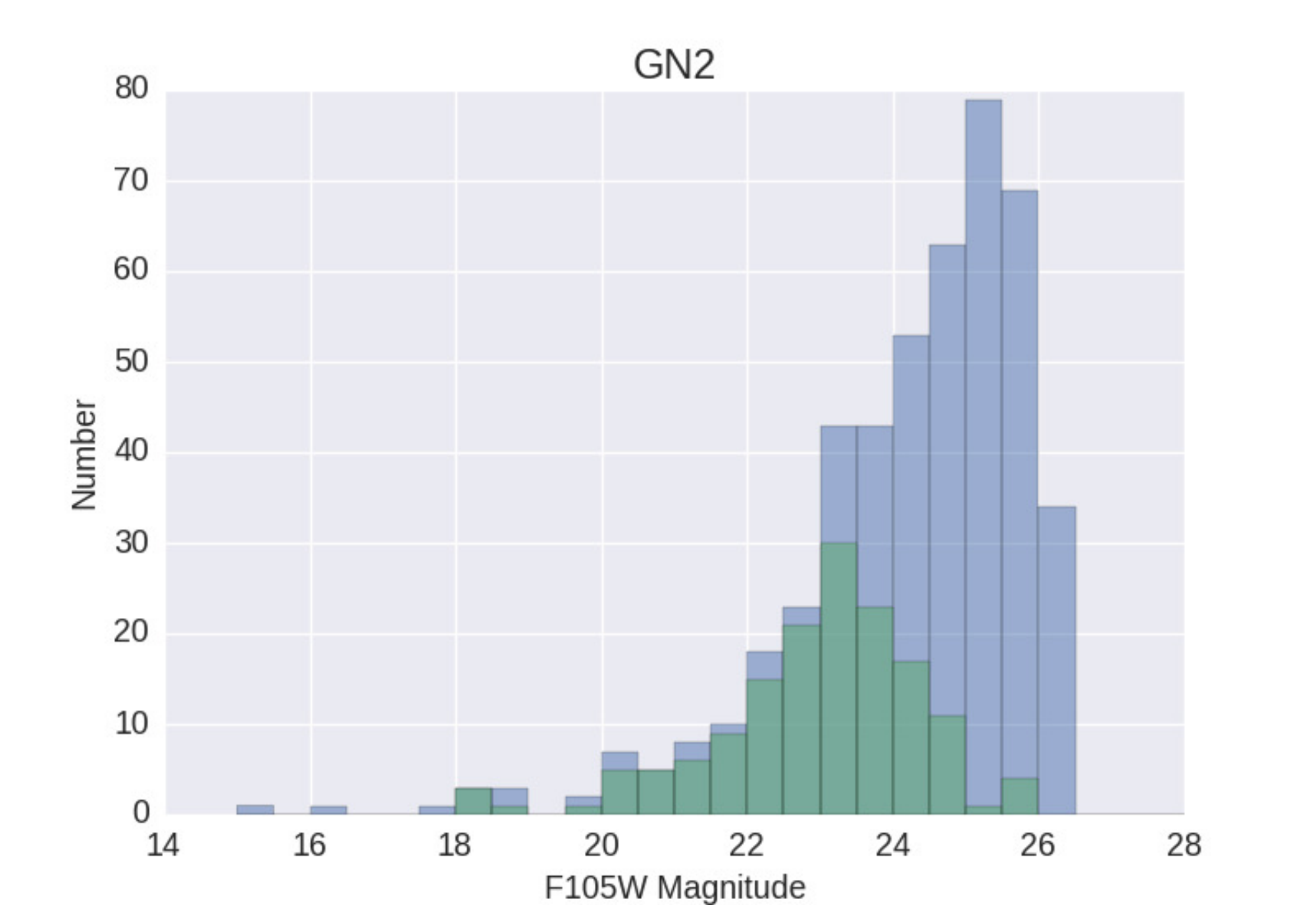} \\
\includegraphics[width=0.5\textwidth]{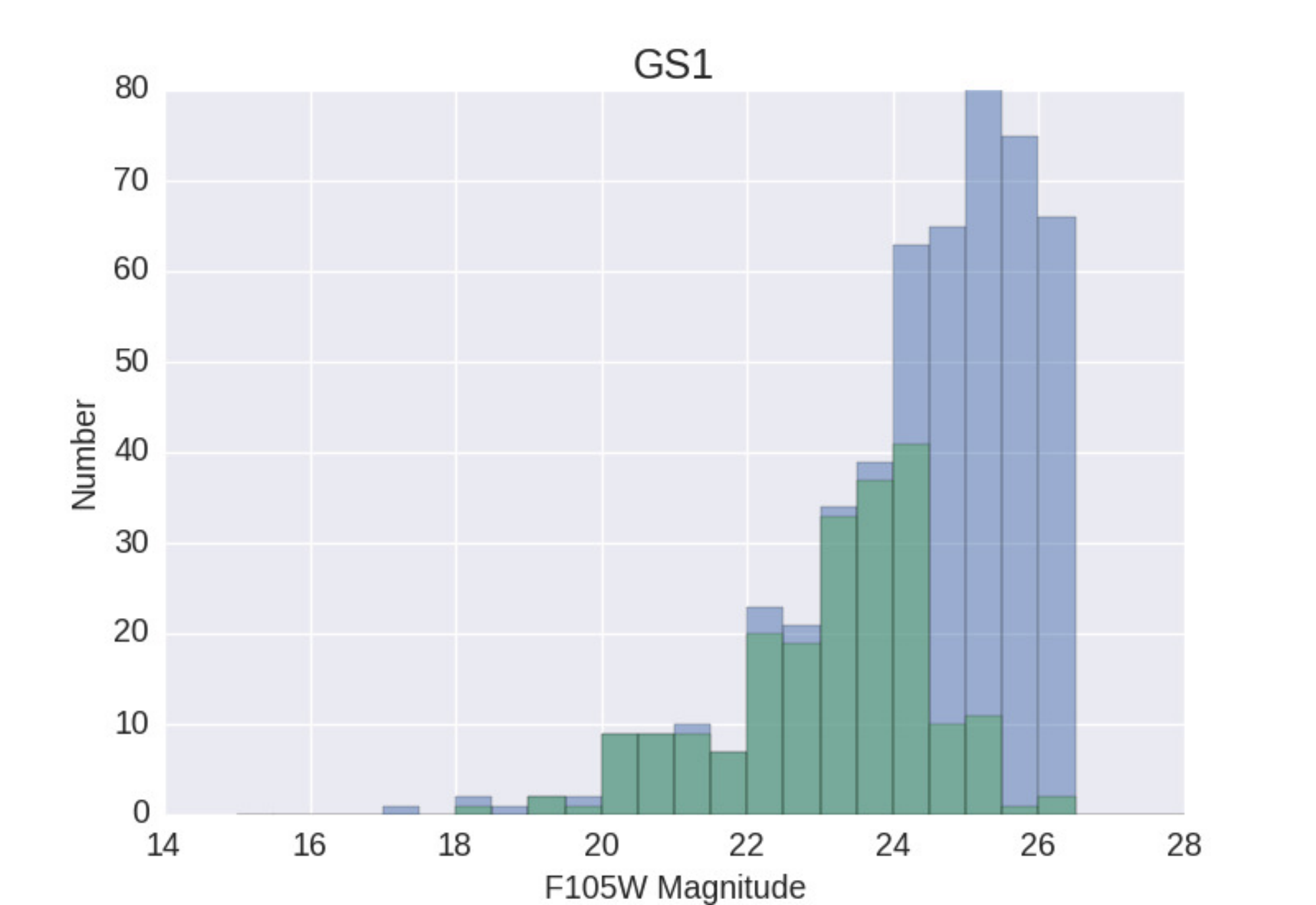} & \includegraphics[width=0.5\textwidth]{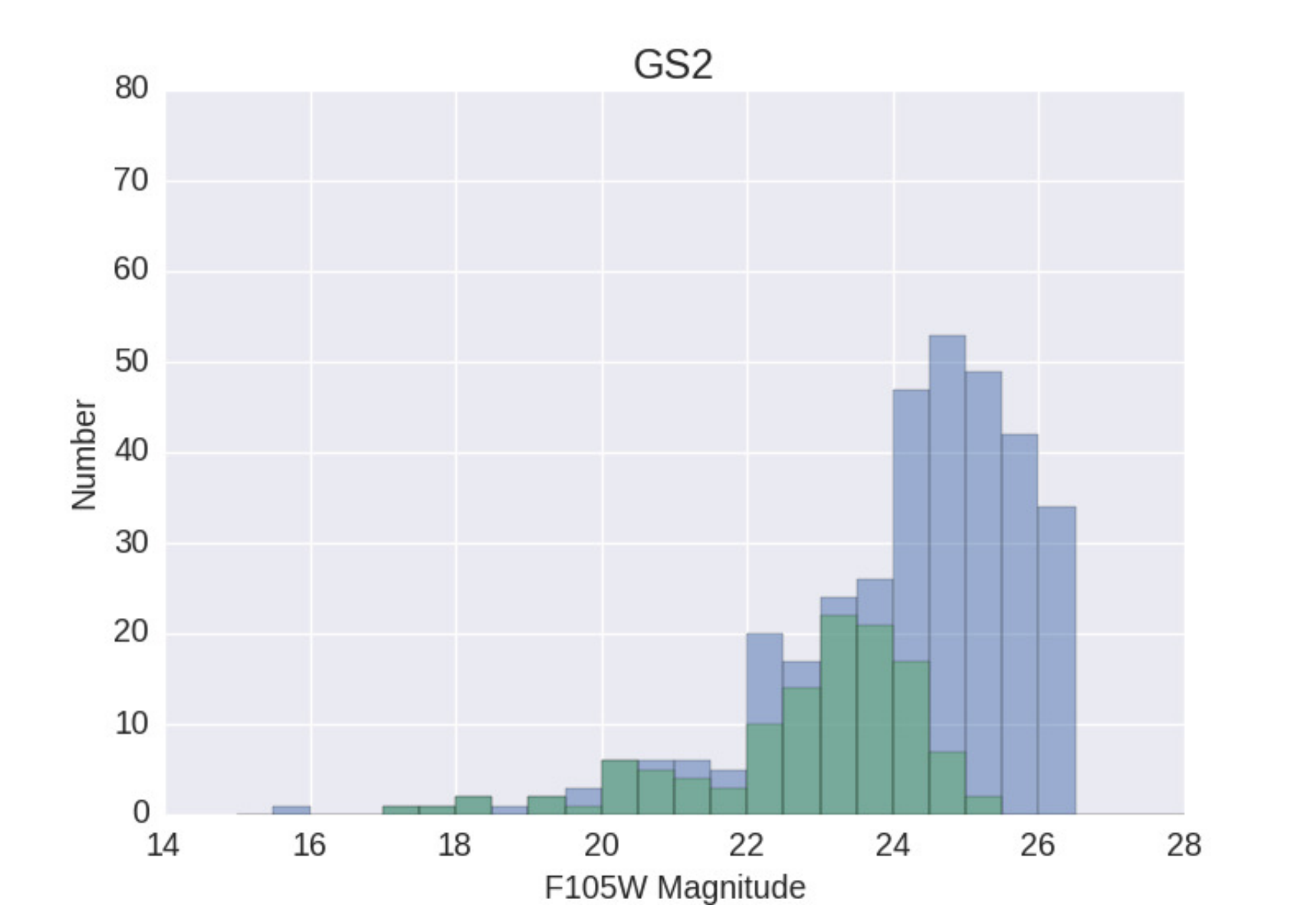} 
\end{tabular}
\caption{These histograms show the F105W magnitude distribution for each field for the total sample of SPZ objects (blue) and the sample of objects with matching spectroscopic redshifts (green). The median values in each sample are (clockwise from top left): GN1, 24.6 mag for SPZs, 23.3 mag for spec-zs; GN2, 24.6 mag for SPZs, 23.2 mag for spec-zs; GS2, 24.5 for SPZs, 23.2 for spec-zs; GS1, 24.8 mag for SPZs, 23.5 mag for spec-zs.}
\end{figure*}

\begin{figure*}
\includegraphics[width=\textwidth]{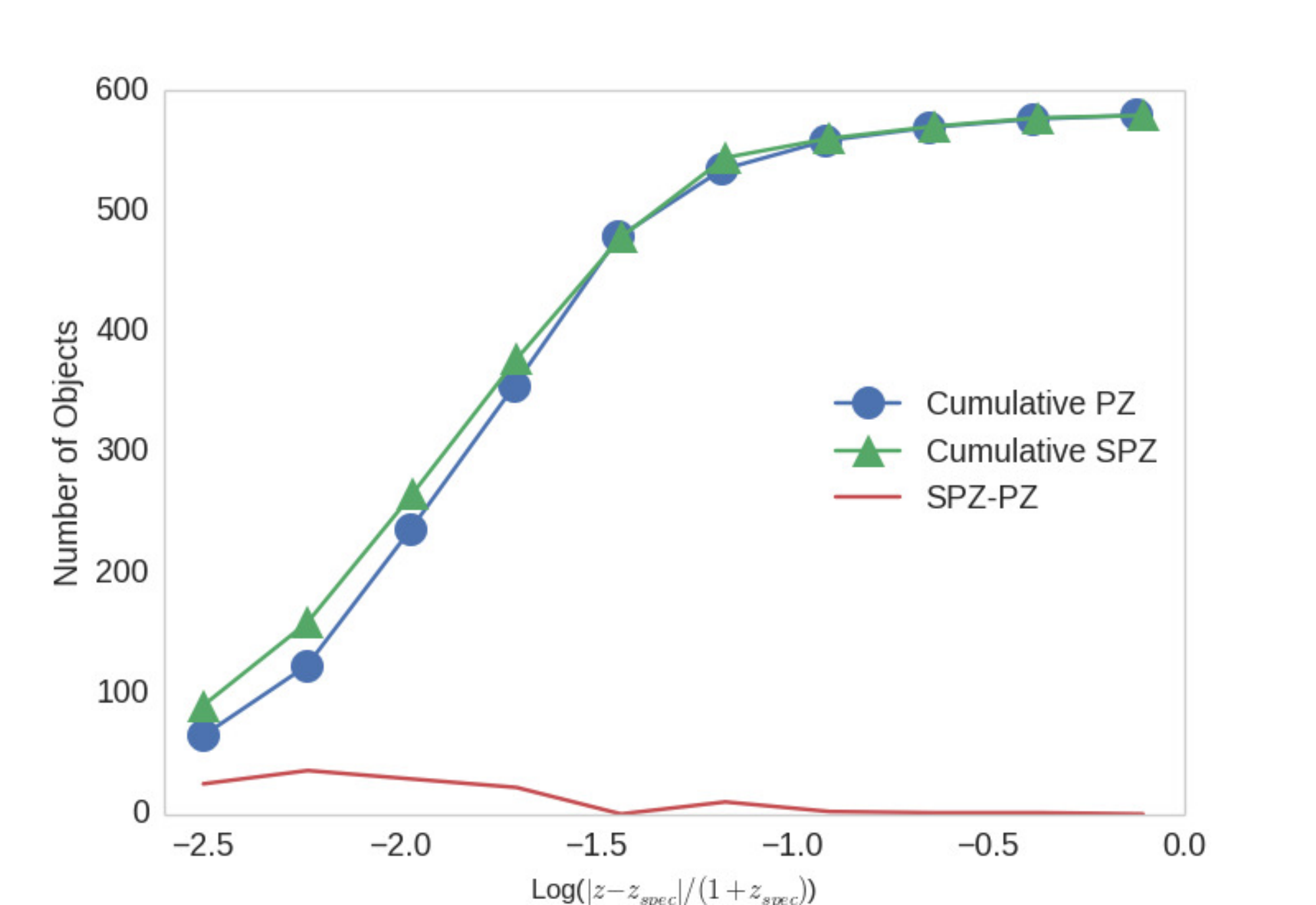}
\caption{A cumulative count of the SPZ-galaxies starting at  $\log (|z - z_{spec}| / (1+z_{spec})) \leq -2.5$. Triangle points (green) show the number of SPZs at a given or greater accuracy. Circle points (blue) show the number of purely photometric redshifts (PZs) at a given or greater accuracy. The red line is the number of SPZs minus the number of PZs, which demonstrates the excess of SPZs in the most accurate bins.}
\end{figure*}

\subsection{SPZ Quality}

\begin{table*}
\begin{center}
\caption{A summary of the SPZ and photo-z quality results for the four FIGS fields.}
\begin{tabular}{ccccccc}
\tableline
Field & N. Spec-z & F105W$^1$ & Med($\Delta z_{SPZ}$)$^2$ & Med($\Delta z_{PZ}$)$^3$ & SPZ Outliers$^4$ & PZ Outliers$^4$ \\
\tableline
GN1 & 200 & 23.3 & 0.019 & 0.026 & 0.07 & 0.08 \\
GN2 & 147 & 23.2 & 0.024 & 0.028 & 0.09 & 0.10 \\
GS1 & 131 & 23.5 & 0.023 & 0.029 & 0.09 & 0.10 \\
GS2 & 101 & 23.2 & 0.027 & 0.031 & 0.15 & 0.16 \\
\tableline
\end{tabular}
\begin{tablenotes}
  \item[]$\rm ^1$ The median F105W magnitude of the SPZ-spectroscopic comparison sample.
  \item[]$\rm ^2$ The quantity described is the median value of $(z_{SPZ} - z_{spec})/(1 + z_{spec})$ for the field.
  \item[]$\rm ^3$ The quantity described is the median value of $(z_{PZ} - z_{spec})/(1 + z_{spec})$ for the field.
  \item[]$\rm ^4$ This refers to the fraction of objects for which the fits are catastrophic failures, meaning $|(z_{SPZ} - z_{spec})/(1 + z_{spec})| > 0.1$.
\end{tablenotes}
\end{center}
\end{table*}

\begin{figure}
\plotone{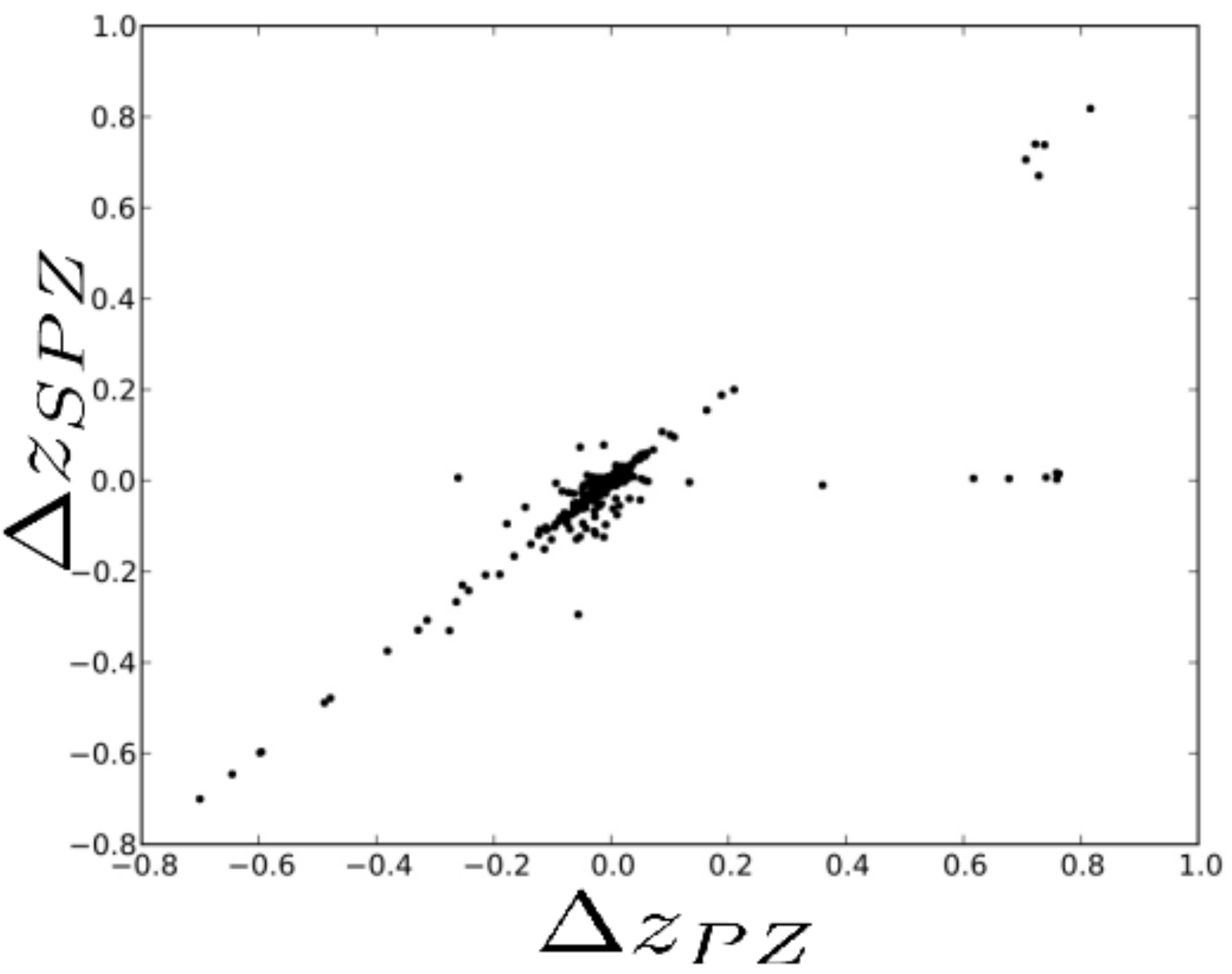}
\caption{A plot of the SPZ difference $\Delta_{SPZ} = (z_{SPZ} - z_{spec})/(1+z_{spec})$, vs the photo-z difference $\Delta_{PZ} = (z_{PZ} - z_{spec})/(1+z_{spec})$. This illustrates the cases where the method of redshift calculation can make a significant change. One can see a number of objects where the photometric redshift produces a $D_{PZ} > 0.1$, a catastrophic failure, while the $D_{SPZ}$ is quite low.}
\end{figure}


\begin{figure*}
\begin{tabular}{cc}    
\includegraphics[width=0.5\textwidth]{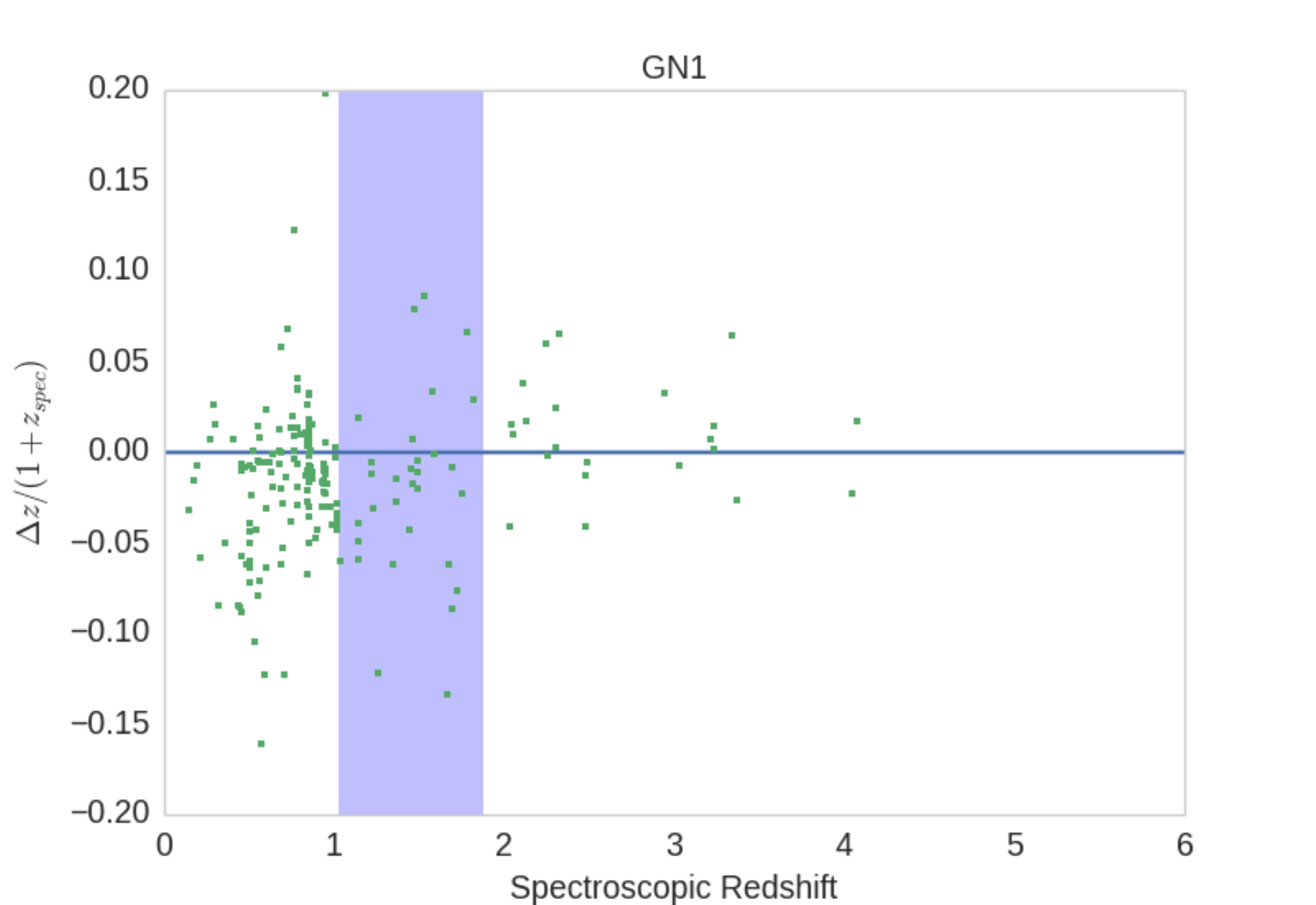} & \includegraphics[width=0.5\textwidth]{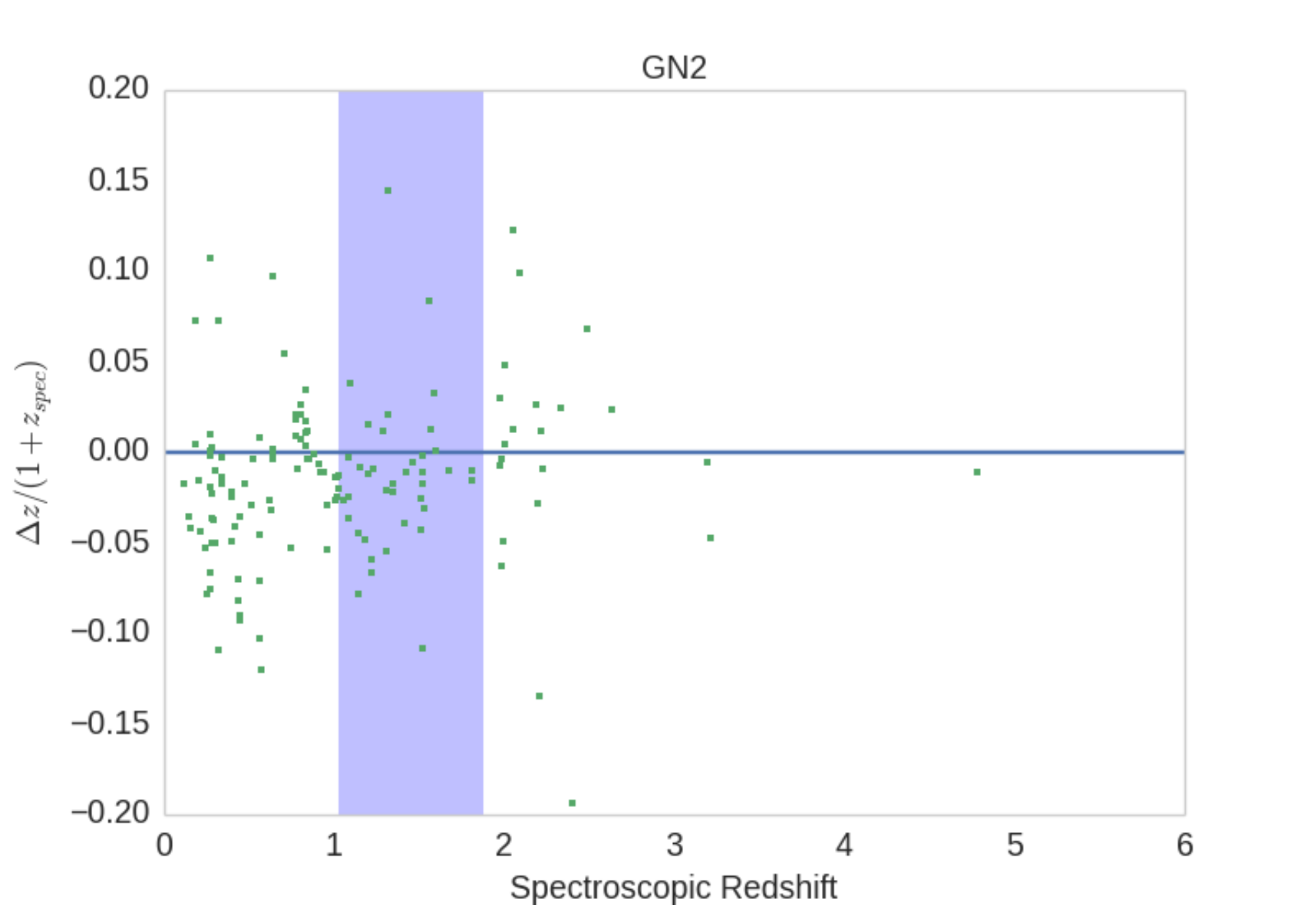} \\
\includegraphics[width=0.5\textwidth]{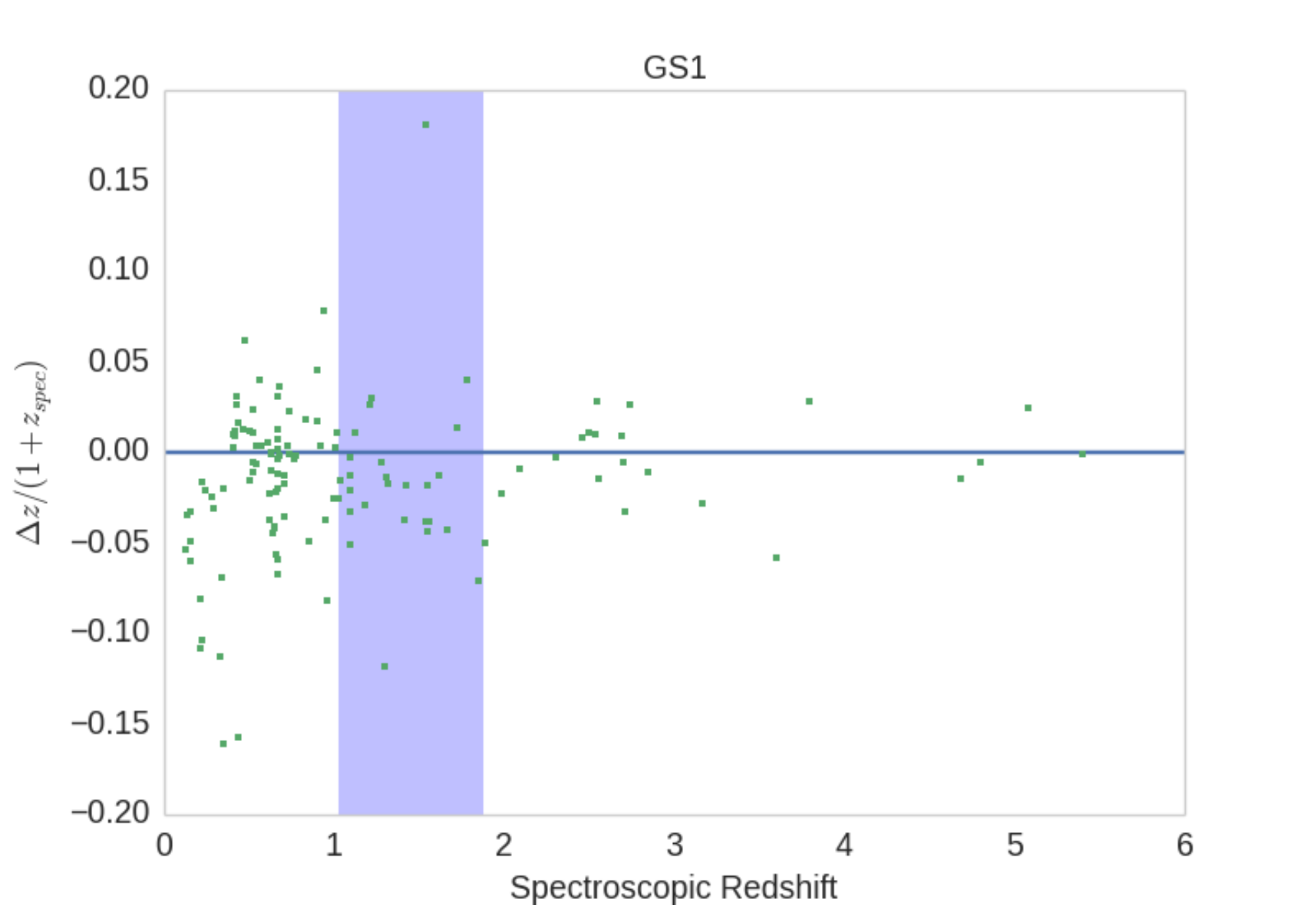} & \includegraphics[width=0.5\textwidth]{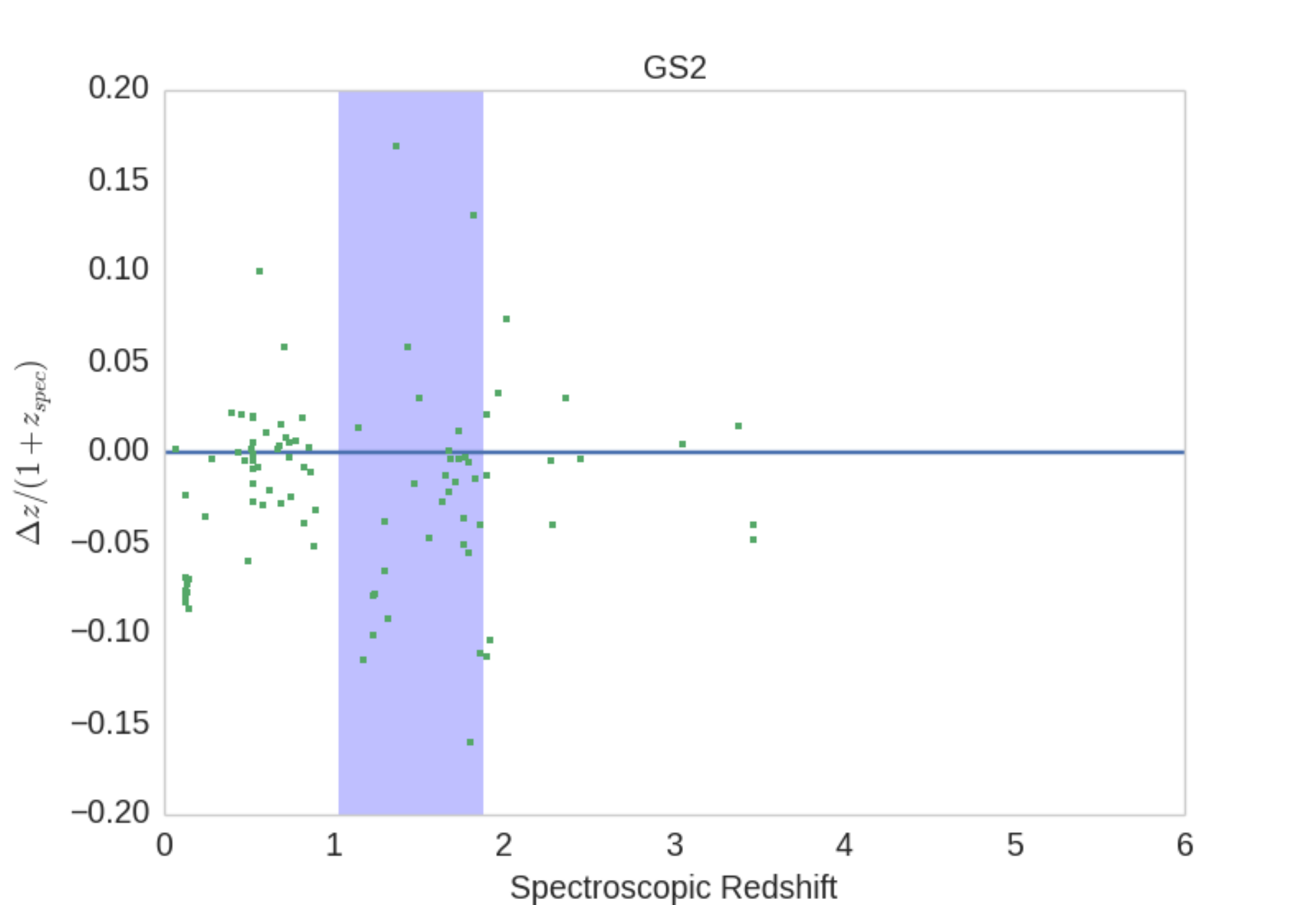} 
\end{tabular}
\caption{The SPZ $\Delta z_{SPZ} = (z_{SPZ} - z_{spec})/(1 + z_{spec})$ versus the known spectroscopic redshifts for each field, clockwise from top left GN1, GN2, GS2, GS1. The blue shaded region reflects the redshift range in which the 4000 \AA\ break falls within the grism wavelength coverage.}
\end{figure*}

\begin{figure*}
\begin{tabular}{cc}    
\includegraphics[width=0.5\textwidth]{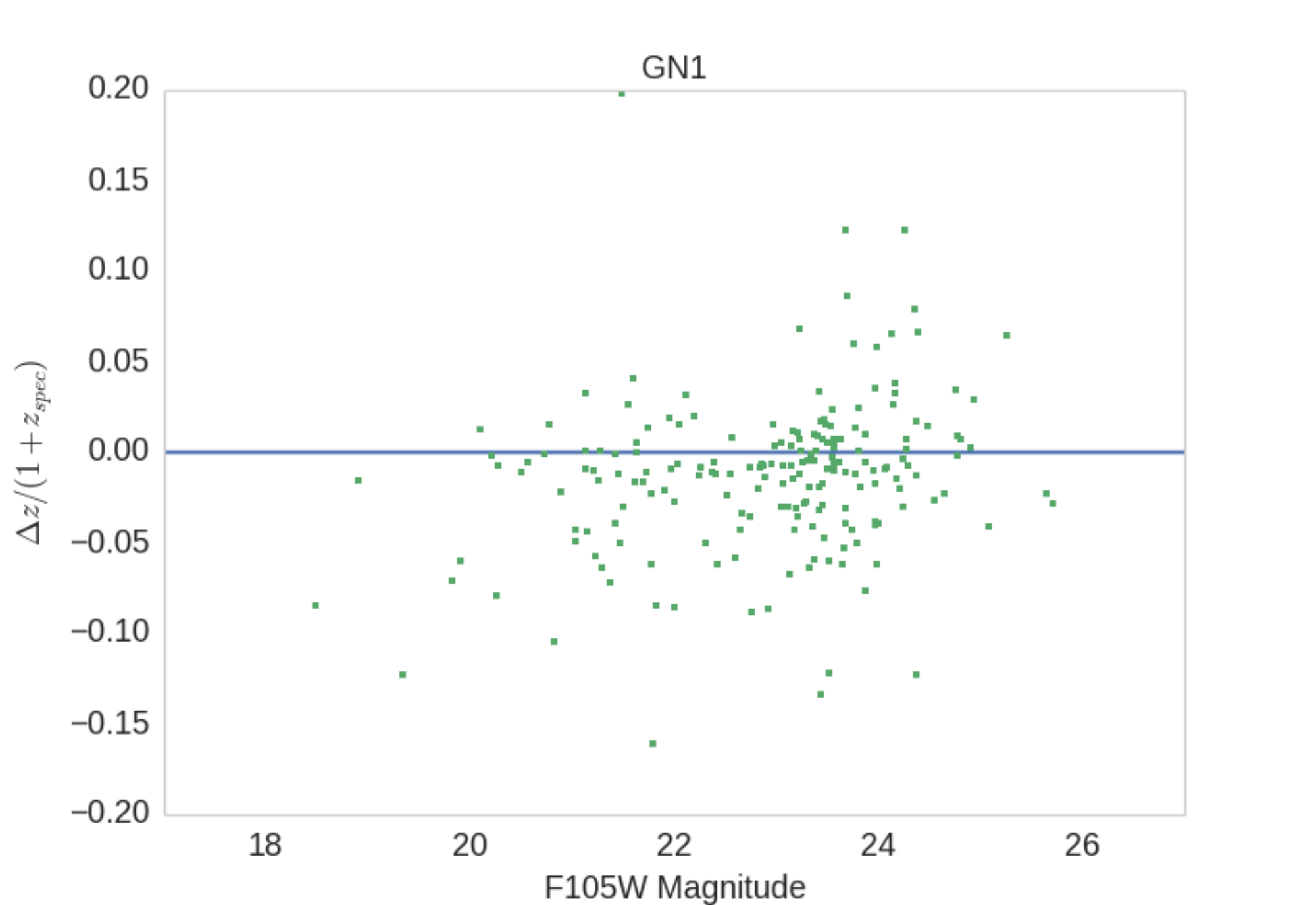} & \includegraphics[width=0.5\textwidth]{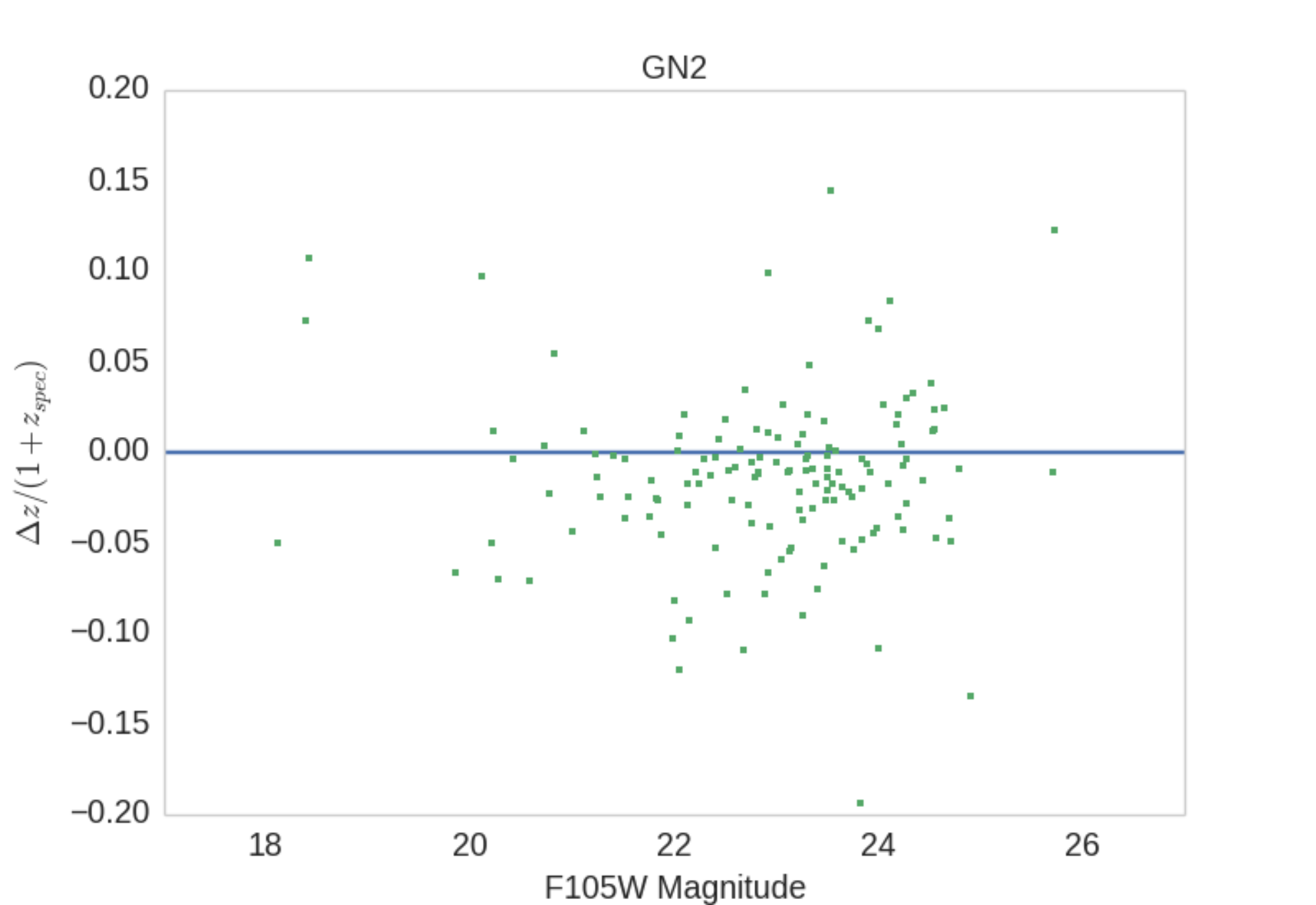} \\
\includegraphics[width=0.5\textwidth]{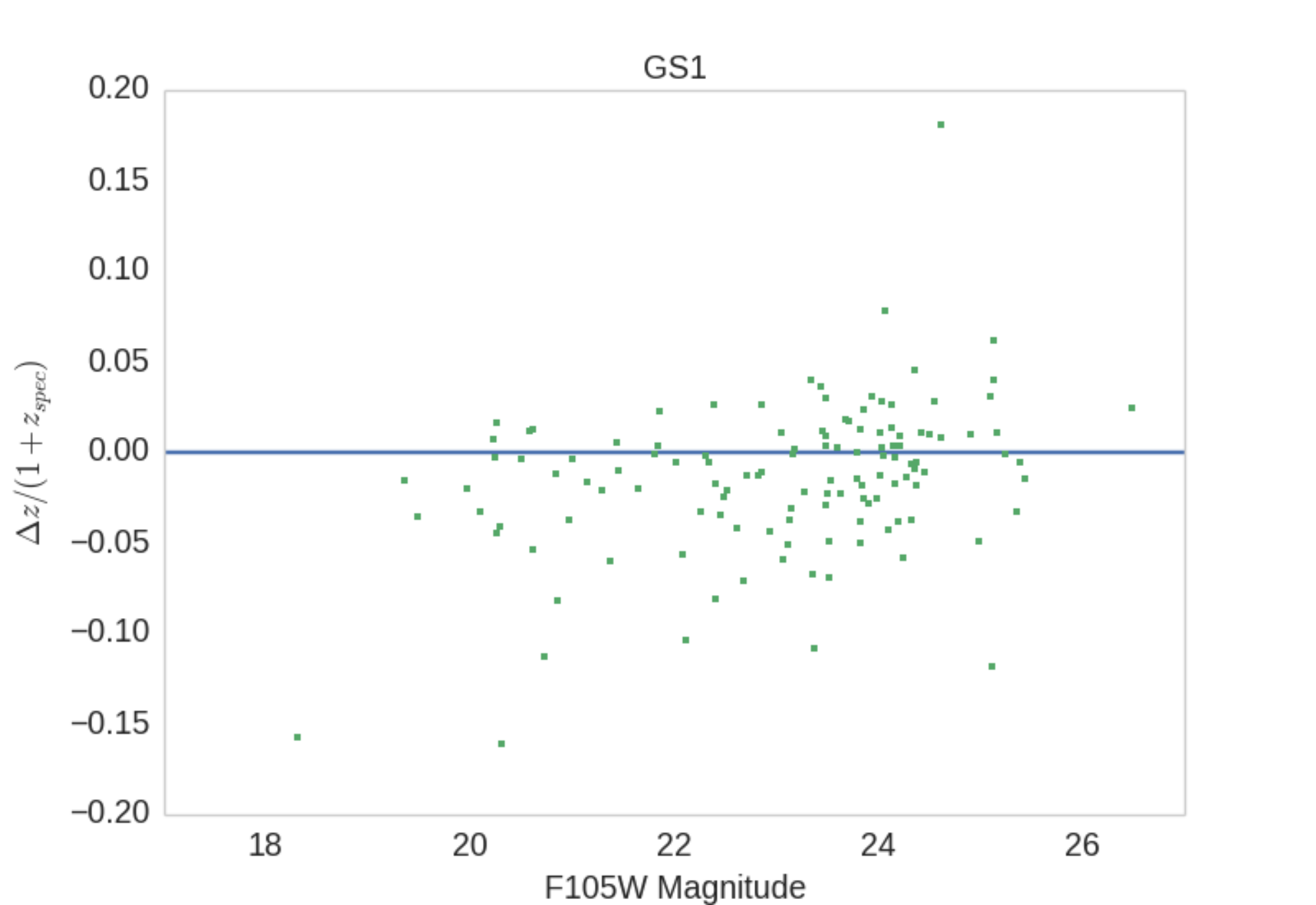} & \includegraphics[width=0.5\textwidth]{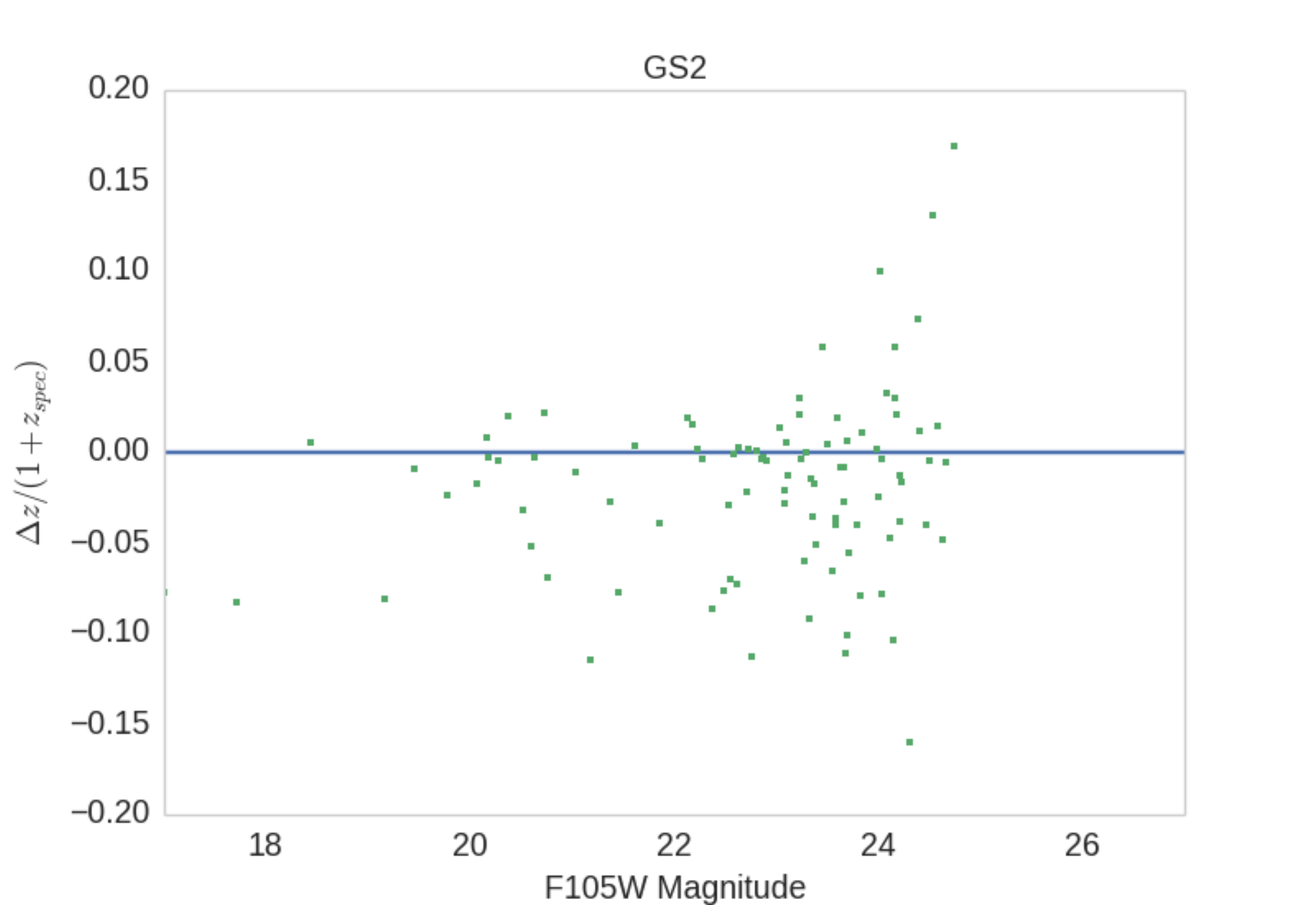} 
\end{tabular}
\caption{The $\Delta z_{SPZ} = (z_{SPZ} - z_{spec})/(1 + z_{spec})$ of SPZ objects versus F105W magnitude for each field, clockwise from top left GN1, GN2, GS2, GS1.}
\end{figure*}

In order to gauge the accuracy of the SPZs compared to photometric redshifts without spectral data (photo-zs), it is helpful to compare the redshifts from both methods to known spectroscopic redshifts (ie, the conventional standard for accurate redshifts). We created a matching set of photometric redshifts for the SPZ objects by simply running the same catalogs through EAZY stripped of their grism measurements, leaving only the broadband photometry as input for the fit.

The SPZ and photo-z catalogs could each be compared to spectroscopic redshifts for the same objects. To find as many matches with confirmed spectroscopic redshifts as possible, we consulted a compilation of public spectroscopic surveys in GOODS-N and CDFS (N. Hathi, private communication). The spectroscopic redshifts in these compilations were assigned quality flags based on the redshift quality indicated in the parent survey. In order to ensure the best possible comparison sample, only the spectroscopic redshifts from the two highest-quality bins were used.

Since the limiting magnitude of the FIGS dataset goes beyond the limits of ground-based spectra, one should expect objects with spectroscopic redshifts in the FIGS fields to be readily detected. Consequently, these objects can be found using simple (RA,DEC) matching. This was done with a separation tolerance of 1 arcsecond to account for offsets in different surveys, though for the vast majority of matches the separation is much smaller. This matched set of spectroscopic redshifts was then assured to provide a high-accuracy comparison for the matched SPZs. The number and magnitudes of the SPZ-spectroscopic comparison sample are given in Table 5, and a comparison of magnitudes between the spectroscopic redshifts and the total sample of SPZ objects are given in Figure 5.

For FIGS objects with existing spectroscopic redshifts of the highest accuracy, we calculate the term:

\begin{equation}
\Delta \mathbf{z_{SPZ}} = \frac{z_{SPZ} - z_{spec}}{1 + z_{spec}}
\end{equation}

\noindent This measures the closeness of each SPZ redshift to the true value given by the spectroscopic redshift. For comparison, we also calculate $\Delta z_{PZ}$ for the purely photometric redshifts. The median $\Delta z_{SPZ}$ and $\Delta z_{PZ}$ for each field is given in Table 5, as well as the outlier rates, defined as the fraction of objects for which $|(z - z_{spec}) / (1 + z_{spec})| > 0.1$. We observe an improvement in $\Delta z_{SPZ}$ over $\Delta z_{PZ}$ from 0.03 to 0.02 in three of the four fields (see Table 5), and an improvement in the outlier rate in all four.

The distribution of redshift accuracy for the entire sample is illustrated in Figures 6 and 7, where the accuracy of a given object's redshift is measured by:

\begin{equation}
  \log \left( \Delta z_{SPZ} \right) = \log \left( \frac{|z_{SPZ} - z_{spec}|}{1 + z_{spec}} \right)
\end{equation}

\noindent
such that more negative results represent redshift fits closer to the spectroscopic redshift. We calculated these for the spectroscopically matched SPZ and photo-z sets, and plotted a histogram of the results in Figure 6. Values of $\log \mathbf{\log(\Delta z_{SPZ})} \leq -2.4$ (which implies $|z_{SPZ} - z_{spec}| \leq 0.004 \cdot (1 + z_{spec})$) were binned together, leading to the larger number of objects seen in the highest-accuracy bin. For the whole sample, the SPZs increase the population of this most-accurate bin by 52\% over photometric redshifts. For F105W $ < 24$ mag, SPZs increase it by 69\%. Figure 7, which plots $\Delta z_{SPZ} $ versus $\Delta z_{PZ}$, provides an alternative comparison of the results, which calls attention to the number of objects that SPZs rescue from catastrophic failure.

The median redshift difference for SPZs is 0.023, and 0.029 for photo-zs. Use of the SPZ method increases the number of objects in the most accurate bin by $\sim 67\%$. Furthermore, one can see that the SPZ method reduces the incidence of catastrophic failure, by reducing the total number of objects for which $\log (\Delta z_{SPZ}) > -1$ from 8\% to 7\% across all four fields. For the subset of objects where F105W $< 24$ mag, the median redshift difference for SPZs is 0.021, and is 0.027 for photo-zs.

Figure 8 shows $\Delta z_{SPZ}$ versus the spectroscopic redshift. The blue shaded region in each plot corresponds to the redshift range in which the 4000 \AA\ break falls within the grism coverage ($z=1.025-1.875$). The improvement in accuracy of SPZs over photometric redshifts in this range is comparable to that of the overall sample, and is larger only in GN2. This could indicate that the addition of grism data can be useful in constraining the SED fit even without the 4000 \AA\ break falling in its range, either by identifying features at other redshifts (eg, emission lines) or by conclusively ruling out the presence of a 4000 \AA\ break where broadband data could not.  This may also be explained by the blue-region objects being fainter: Figure 9 shows that the majority of catastrophic failures occur beyond F105W $> 24$ mag. There are also considerably fewer objects in this range with high-accuracy spectroscopic redshifts compared to lower redshifts. These figures also seem to show a slight systematic offset in $(z - z_{spec}) / (1 + z_{spec})$: the median $(z - z_{spec}) / (1 + z_{spec}) \sim -0.01$ in both SPZs and photo-zs, suggesting a tendency to slightly underestimate the redshift.This could perhaps be explained by the misidentification of the Balmer break (3646 \AA) as the 4000 \AA\ break, or the application of the magnitude prior to the redshift calculation could be causing a slight preference for lower redshifts. The results given in Table 5 reflect the median error without correcting for this bias.




\section{Finding Galaxy Overdensities}

\begin{figure*}
\plotone{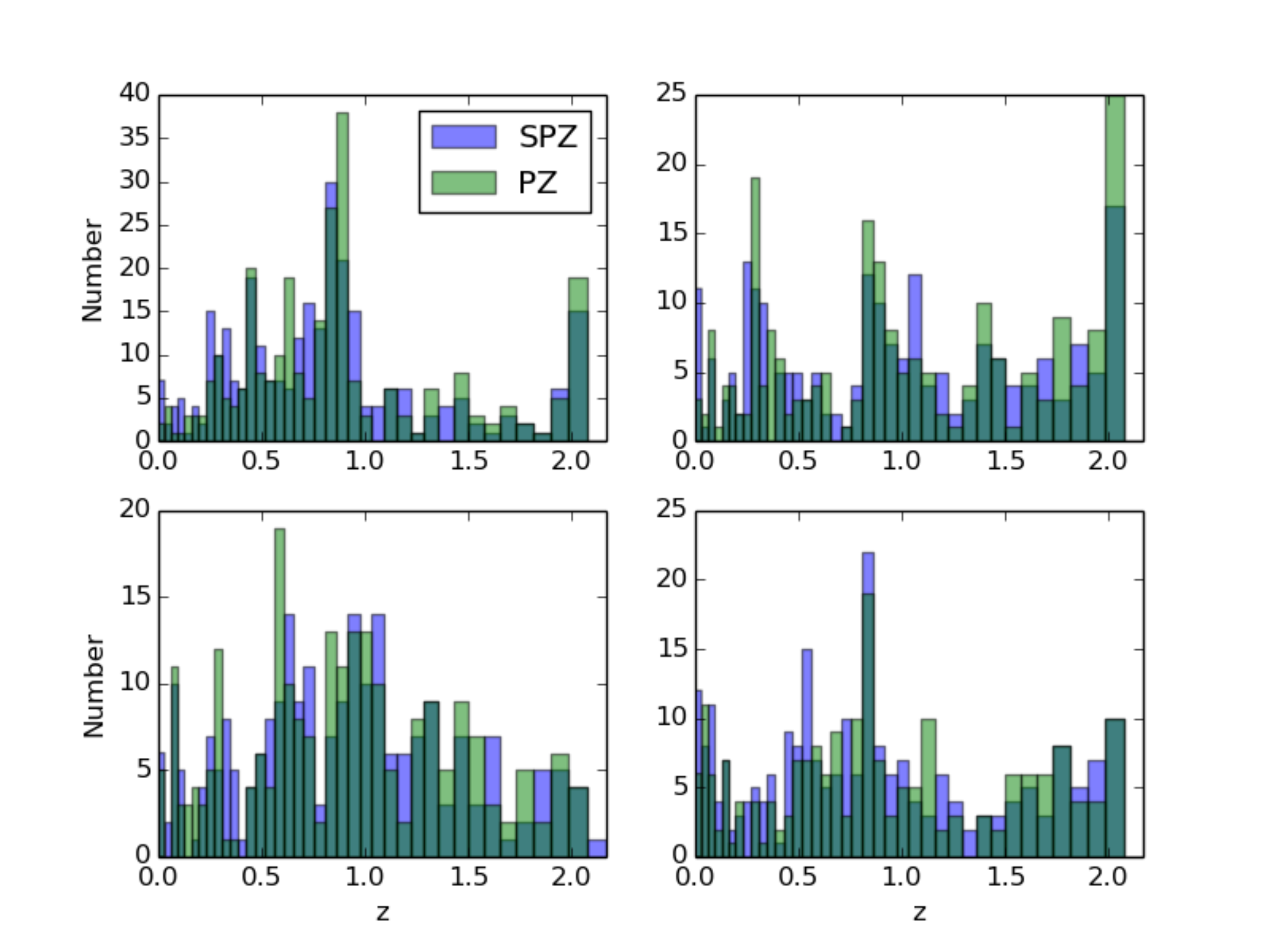}
\caption{A breakdown of the redshift distributions by FIGS field (clockwise from top left: GN1, GN2, GS2, GS1), using both SPZs (blue) and photo-zs (green). The bin widths are given by $\Delta z = 0.03 \cdot (1 + z)$ in order to roughly match the error threshold of the redshifts. The full photo-z dataset could include many more objects than are presented here; this includes only those with a matching SPZ.}
\end{figure*}

After confirming the accuracy of the SPZ set, we performed a pilot study by analyzing the FIGS fields for evidence of significant overdensities in LSS. We began by constructing one-dimensional redshift distributions for each field (Figure 10). This preliminary analysis shows some possible redshift peaks, including one at $z \simeq 0.85$ in GN1. The peak is a bit more dominant in the photometric redshift set at a somewhat higher redshift, but still noticeable in the SPZ. Furthermore, given the lower accuracy and higher outlier rate among photo-zs, peaks in the distribution are more likely to be spurious.

\begin{figure*}
\begin{tabular}{cc}
  \includegraphics[width=0.5\textwidth]{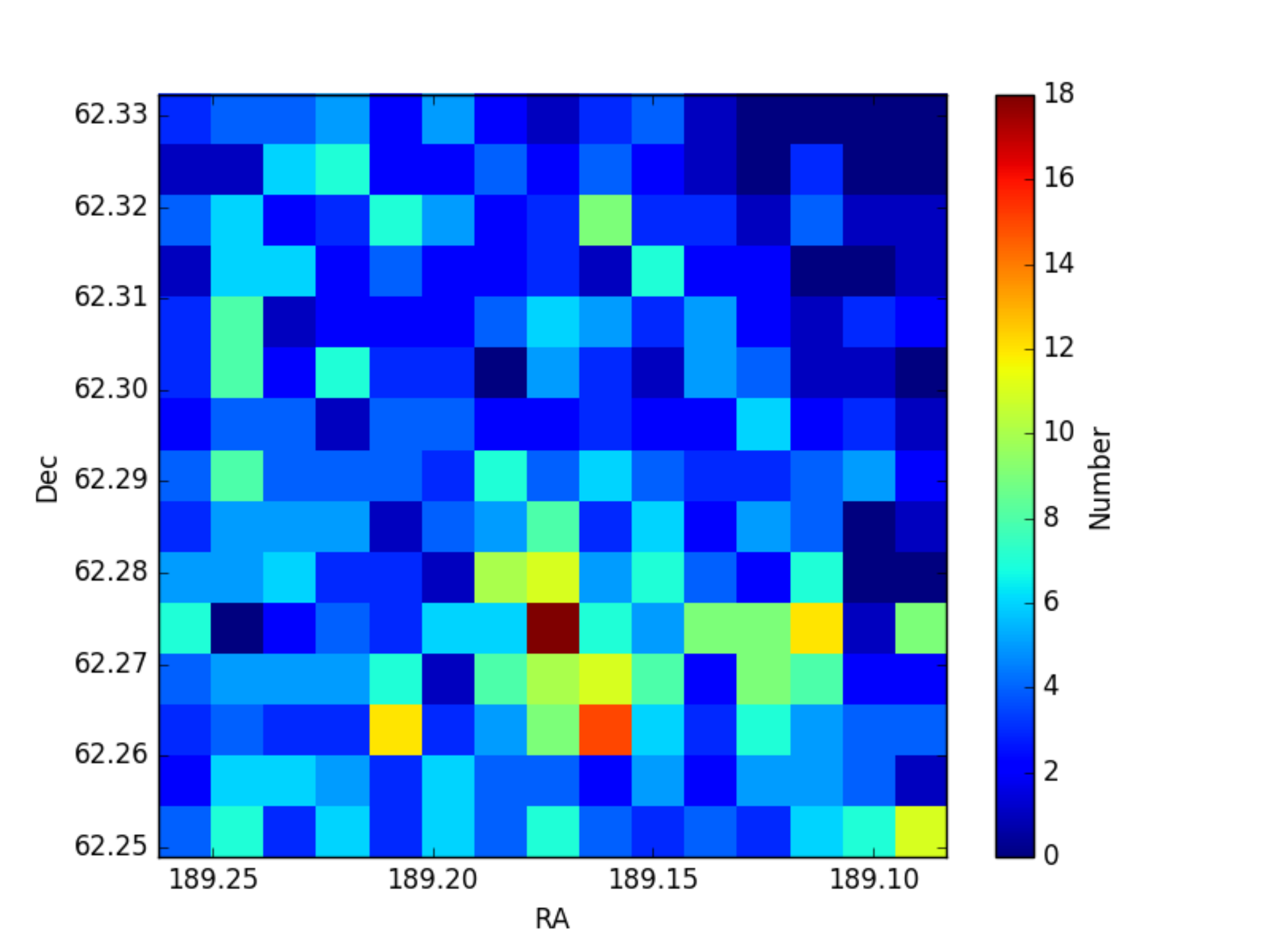} & \includegraphics[width=0.5\textwidth]{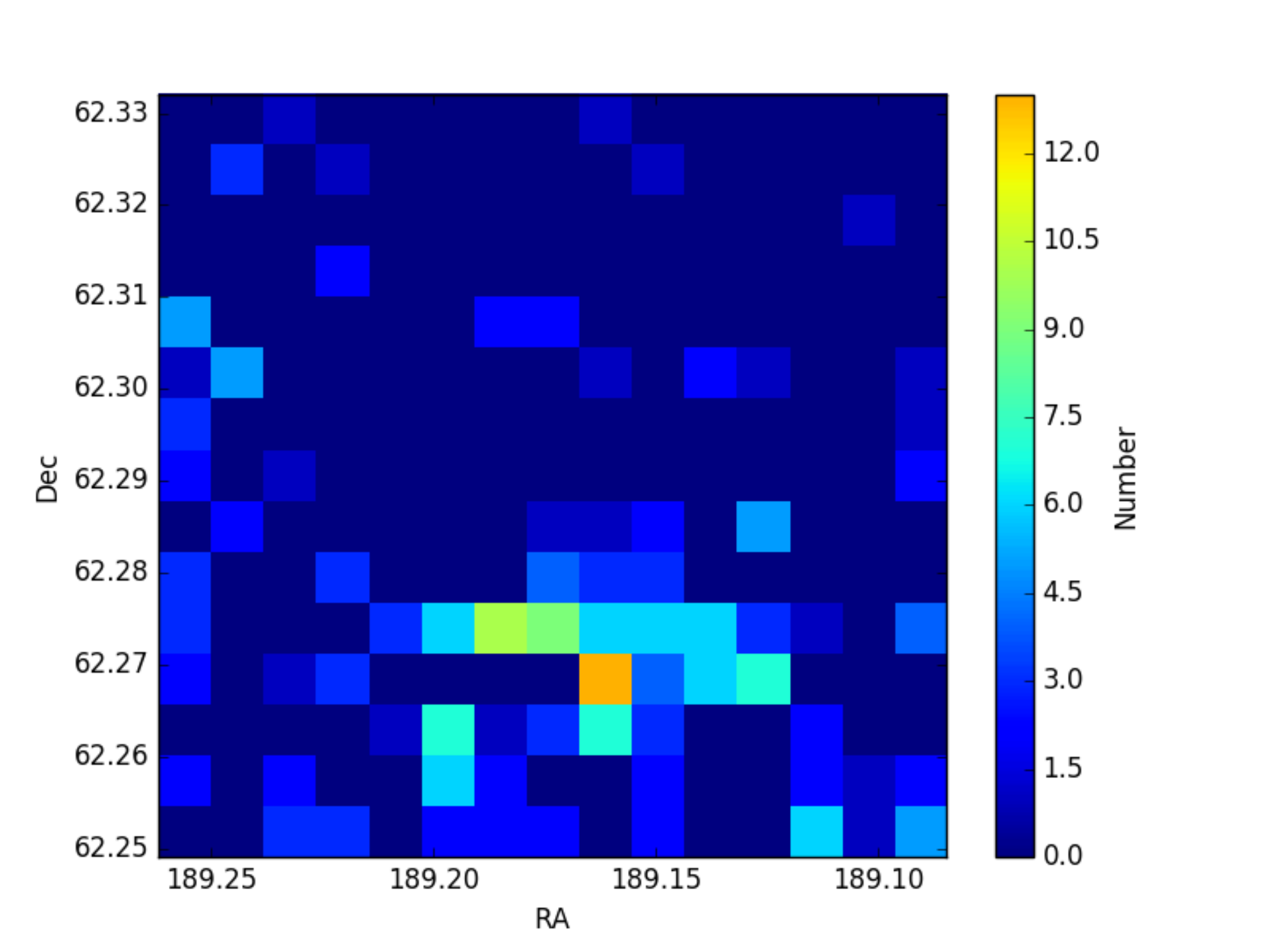}
\end{tabular}
\caption{A 2D histogram of redshift $z \sim 0.85$ objects in GN1 from SPZs (left) and spectroscopic redshifts (right). The color of each square bin scales with the number of objects contained in that angular area. Since the objects have already been selected for a narrow redshift range, correlation and overdensity of objects in this plot indicate a spatial correlation. The mean number of objects per bin in GN1 in the SPZ plot is $\sim 4$.}
\end{figure*}

\subsection{Confirmation of a Previously Known Overdensity at $z=0.85$}

To see if there was a matching angular overdensity, we plotted the J2000.0 (RA, DEC) positions of the objects in this peak redshift bin in a two-dimensional histogram (Figure 11, left). This shows several points with a high concentration of objects, the peak of which has a number of sources $\sim 4$ times the mean in GN1.

\begin{table*}
\begin{center}
  \caption{Potential overdensities identified through the nearest-neighbor method.}
    \begin{tabular}{cccccccccc}
\tableline
ID & RA & Dec & $z_1$ & $\mathcal{M}$\footnotemark[1] & $\mathcal{S}$\footnotemark[2] & $z_2$ & $\mathcal{M}_2$ & $\mathcal{S}_2$ & N. Galaxies\\
\tableline
GN1-0.2 & 189.211302 & +62.303195 & 0.2-0.3 & 11.90 & 47.10 & - & - & -  & 43\\
GN1-0.3 & 189.148053 & +62.292436 & 0.25-0.35 & 16.46 & 21.28 & 0.315-0.354 & 10.05 & 10.18 & 10 \\
GN1-0.4 & 189.202382 & +62.276415 & 0.4-0.5 & 10.54 & 13.68 & 0.435-0.478 & 5.89 & 10.30 & 26 \\
GN1-0.6 & 189.141025 & +62.290214 & - & - & - & 0.645-0.694 & 20.43 & 16.15 & 24 \\
GN1-0.7 & 189.177515 & +62.272790 & - & - & - & 0.735-0.787 & 17.01 & 18.10 & 29 \\
GN1-0.8\footnotemark[3] & 189.162108 & +62.281327 & 0.8-0.9 & 25.16 & 25.09 & 0.825-0.880 & 28.23 & 24.41 & 28 \\
GN1-1.2 & 189.191840 & +62.281093 & 1.25-1.35 & 15.87 & 10.28 & 1.290-1.359 & 13.31 & 21.49 & 7 \\
GN1-1.6 & 189.149675 & +62.287408 & 1.6-1.7 & 10.66 & 21.250 & 1.680-1.760 & 7.40 & 12.86 & 13\\
GN1-1.9 & 189.203733 & +62.277585 & 1.95-2.05 & 12.97 & 50.35 & 1.965-2.054 & 17.51 & 15.60 & 24 \\
GN1-3.2 & 189.185353 & +62.280742 & - & - & - & 3.21-3.336 & 27.43 & 18.41 & 13 \\
\tableline
GN2-0.2 & 189.358060 & +62.290507 & 0.2-0.3 & 13.80 & 41.35 & 0.255-0.293 & 8.29 & 64.43 & 31\\
GN2-0.5 & 189.357597 & +62.310505 & 0.5-0.6 & 10.02 & 11.59 & 0.540-0.586 & 5.81 & 12.79 & 16 \\
GN2-0.9 & 189.376327 & +62.290928 & - & - & - & 0.915-0.972 & 11.05 & 15.04 & 18 \\
\tableline
GS1-0.1 & 53.185189 & -27.791704 & 0.1-0.2 & 10.76 & 12.91 & 0.105-0.138 & 10.67 & 26.33 & 15 \\
GS1-0.2 & 53.166696	& -27.787796 & 0.25-0.35 & 10.85 & 27.49 & 0.270-0.308 & 12.72 & 26.22 & 29 \\
GS1-0.5 & 53.170932 & -27.794401 & 0.5-0.6 & 12.32 & 10.74 & - & - & - & 23 \\
GS1-0.7 & 53.161736	& -27.789378 & - & - & - & 0.765-0.818 & 10.07 & 15.83 & 29 \\
GS1-0.9 & 53.158430 & -27.787889 & 0.9-1.0 & 11.52 & 47.80 & 0.900-0.957 & 6.04 & 16.14 & 48 \\
GS1-1.8 & 53.153574 & -27.777564 & 1.8-1.9 & 18.20 & 41.65 & 1.815-1.899 & 20.41 & 12.83 & 22\\
\tableline
GS2-0.0 & 53.276137 & -27.856452 & 0.05-0.15 & 48.87 & 160.61 & 0.075-0.107 & 100.00 & 378.80 & 17 \\
GS2-0.1 & 53.275942 & -27.861384 & 0.1-0.2 & 15.03 & 11.41 & - & - & - & 23 \\
GS2-0.7 & 53.282187 & -27.861471 & 0.7-0.8 & 12.11 & 23.68 & 0.780-0.833 & 15.59 & 32.57 & 28 \\
GS2-1.6 & 53.288724 & -27.859740 & - & - & - & 1.665-1.745 & 16.10 & 17.19 & 20 \\
GS2-1.7 & 53.286967 & -27.859913 & 1.7-1.8 & 11.85 & 14.58 & 1.710-1.807 & 13.78 & 13.60 & 19 \\
\tableline
\end{tabular}
\begin{tablenotes}
\item The first set of ($z$, $\mathcal{M}$, $\mathcal{S}$) was derived from the nearest-neighbor search with $\Delta z = 0.1$. The second set was derived with $\Delta z = 0.03 \cdot (1 + z_{prev})$. $\mathcal{M}$ and $\mathcal{S}$ are significance measures detailed in \S5.2.
\item[]$\rm ^a$ This number is given by the ratio of the peak nearest-neighbor density in the redshift slice to the median density for the slice.
\item[]$\rm ^b$ This number is given by the ratio of the peak nearest-neighbor density in the redshift slice to the standard deviation in the densities of the adjacent redshift slices.
  \item[]$\rm ^c$ Cluster confirmed in \citet{daw01}
\end{tablenotes}
\end{center}
\end{table*}

\begin{table*}
\begin{center}
\caption{Overdensity Search Comparison}
\begin{tabular}{ccccccc}
\tableline
ID & $z_{SPZ}$ & $\mathcal{M}_{SPZ}$ & $\mathcal{S}_{SPZ}$ & $z_{PZ}$ & $\mathcal{M}_{PZ}$ & $\mathcal{S}_{PZ}$ \\
\tableline
GN1-0.3 & 0.315-0.354 & 10.05 & 10.18 & - & - & - \\
GN1-0.4 & 0.435-0.478 & 5.89 & 10.30 & - & - & - \\
GN1-0.5 & - & - & - & 0.510-0.555 & 10.79 & 10.56 \\
GN1-0.6 & 0.645-0.694 & 20.43 & 16.15 & 0.630-0.679 & 21.75 & 19.80 \\
GN1-0.7 & 0.735-0.787 & 17.01 & 18.10 & 0.750-0.803 & 24.95 & 12.88 \\
GN1-0.8 & 0.825-0.880 & 28.23 & 24.41 & 0.870-0.926 & 15.11 & 13.18 \\
GN1-1.2 & 1.290-1.359 & 13.31 & 21.49 & 1.290-1.359 & 10.56 & 13.46 \\
GN1-1.6 & 1.680-1.760 & 7.40 & 12.86 & - & - & - \\
GN1-1.9 & 1.965-2.054 & 17.51 & 15.60 & - & - & - \\
GN1-3.2 & 3.210-3.336 & 27.43 & 18.41 & - & - & - \\
\tableline
GN2-0.2 & 0.255-0.293 & 8.29 & 64.43 & 0.255-0.293 & 7.32 & 27.85 \\
GN2-0.5 & 0.540-0.586 & 5.81 & 12.79 & - & - & - \\
GN2-0.9 & 0.915-0.972 & 11.05 & 15.04 & 0.930-0.988 & 7.04 & 10.81 \\
\tableline
GS1-0.1 & 0.105-0.293 & 10.67 & 26.33 & - & - & - \\
GS1-0.2 & 0.270-0.308 & 12.72 & 26.22 & 0.255-0.293 & 4.71 & 16.67 \\
GS1-0.5 & - & - & - & 0.495-0.54 & 18.65 & 34.87 \\
GS1-0.7 & 0.765-0.818 & 10.07 & 15.83 & 0.765-0.818 & 4.78 & 10.60 \\
GS1-0.9 & 0.900-0.957 & 6.04 & 16.14 & - & - & - \\
GS1-1.8 & 1.815-1.899 & 20.41 & 12.83 & 1.83-1.915 & 5.77 & 14.05 \\
\tableline
GS2-0.0 & 0.075-0.107 & 100.00 & 378.80 & 0.06-0.092 & 10.49 & 25.48 \\
GS2-0.7 & 0.780-0.833 & 15.59 & 32.57 & 0.795-0.849 & 13.25 & 11.82 \\
GS2-1.6 & 1.665-1.745 & 16.10 & 17.19 & 1.68-1.76 & 21.01 & 13.81 \\
GS2-1.7 & 1.710-1.807 & 13.78 & 13.60 & 1.725-1.807 & 19.90 & 14.75 \\
\tableline
\end{tabular}
\begin{tablenotes}
\item $\mathcal{M}$ and $\mathcal{S}$ are significance measures detailed in \S5.2
\end{tablenotes}
\end{center}
\end{table*}


\begin{figure*}
  \includegraphics{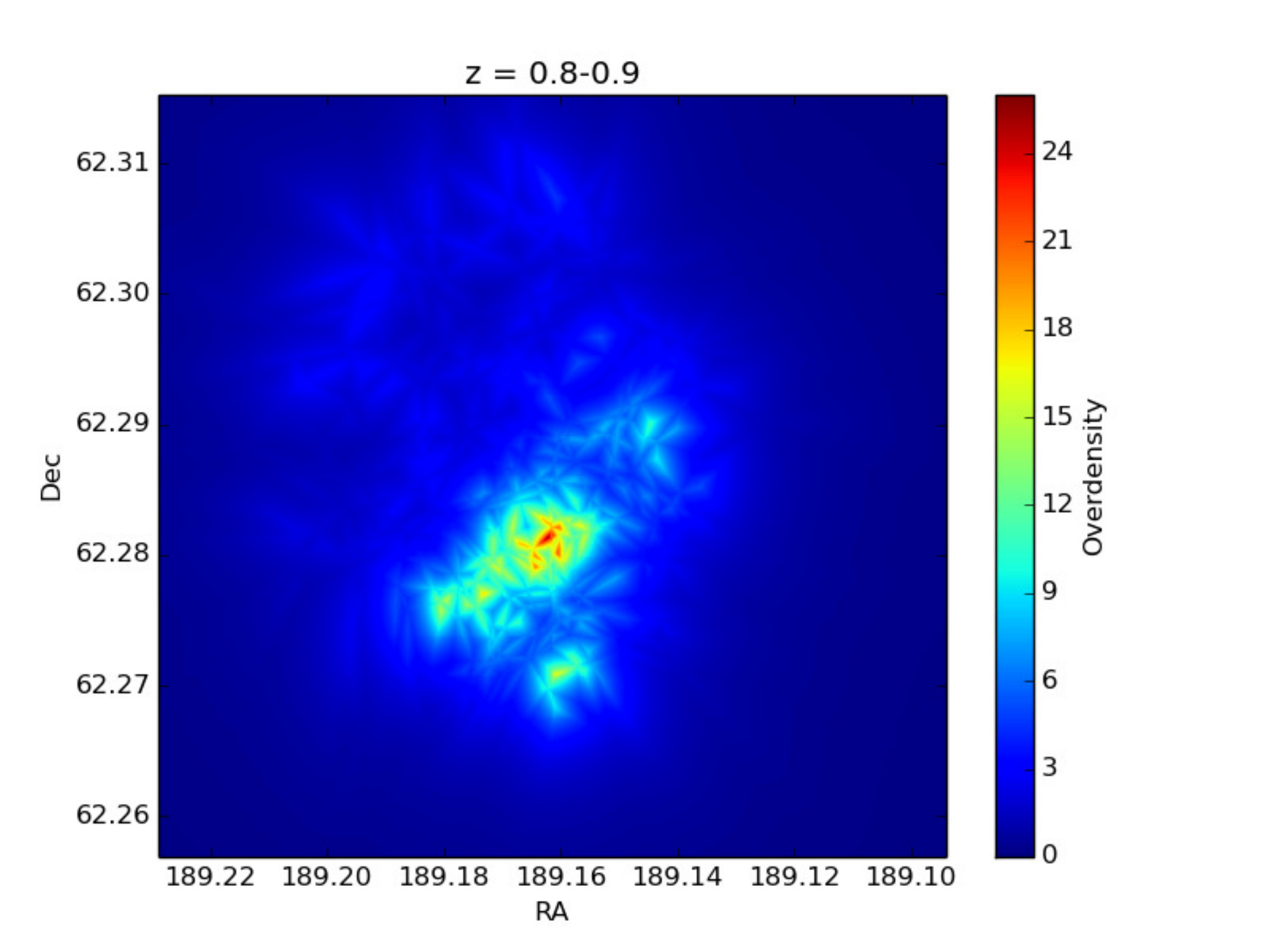}
  \caption{A 7th-nearest-neighbor density map for GN1 in the redshift slice $z=0.8$ to $z=0.9$. The color corresponds to the Overdensity factor, which is the density at a given point normalized by the median density of the whole field. The location of the peak overdensity corresponds to a serendipitously spectroscopically identified galaxy cluster at $z=0.85$.}
\end{figure*}


The same process was repeated with the spectroscopic redshift dataset, which shows an overdensity in the same region. To assess this overdensity, we applied a method used for the identification of a candidate cluster in Z-FOURGE \citep{spi12}. We used SPZs to construct a 7th-nearest-neighbor density distribution for the $z=0.8 - 0.9$ redshift slice in GN1 (Figure 12). This was accomplished by constructing a 500x500 grid of points across the whole GN1 field. For each point, the number density of nearby objects was determined by:

\begin{equation}
  n = \frac{N}{\pi r_7^2}
\end{equation}
  
\noindent where $r_7$ is the distance from the point to its seventh-nearest neighbor and $N=7$ is the number of objects in the redshift slice within the distance $r_7$. Once this density is calculated for each point in the field, the mean nearest-neighbor density for the slice is determined and used to scale the densities. Spitler and others have tested nearest-neighbor results for values of $N$ ranging from 5-9, and find little change in the significance of cluster detection \citep{pap10}. We performed this analysis for varying values of $N$ as well, and find the same result (see \S 5.2).

The coordinates and redshift of this overdensity correspond to a $z=0.85$ galaxy cluster serendipitously identified with spectroscopic redshifts by \citet{daw01}. However, the nearest-neighbor density plot shown in Figure 12 indicates some possible substructures within the overall cluster, which is difficult to identify with the smaller spectroscopic sample alone. 

\subsection{Systematic Search for LSS} 

Having verified the viability of the method by recovering the $z=0.85$ cluster, we applied the same nearest neighbor calculation to the rest of the FIGS dataset in slices of $\Delta z = 0.1$. First, we checked the appropriateness of using $N = 7$ for a nearest-neighbor radius $r_{N}$ by recalculating the density map for the same slice with the value of $N$ varying from 5 to 10. For values $N > 7$, the overdensity is still present, though the significance is diminished with respect to the field background, peaking at 6-7 times the mean density rather than 14.3. For $N < 7$, the significance of the $z = 0.85$ cluster remains at a level comparable to $N = 7$, but other regions in GN1 where there is no spectroscopically confirmed overdensity increased to a significance unsupported by the spectroscopic coverage. Thus, we settled on $N = 7$, as it demonstrated the best confirmation of the existing overdensity.

We applied the $n_7$ calculation to each field in slices of $\Delta z = 0.1$ from $z=0$ to $z=6$ (the maximum redshift we allowed in the EAZY calculation). If a slice contained too few objects to perform the calculation, it was skipped (as was the case for many of the high-z slices). In order to avoid boundary misses, where an overdensity would be missed if its mean redshift were at the boundary between two $\Delta z$ steps, we iterated in steps of $0.5 \cdot \Delta z$.

For these slices, we applied two different measures of overdensity significance. First, we normalized each point in the density grid by the median nearest neighbor density for that redshift slice. This is superior to normalizing by the mean, since the value of the mean will be biased toward a high density peak if one exists. For each slice, we recorded the peak median-normalized density (called $\mathcal{M}$). This checks the significance of an overdense region relative to the density of the entire field at a given redshift range, but may underestimate significance if the angular size of the structure is large relative to the size of the whole field. For the second method, which is based on the method used by Spitler, we calculated the standard deviation in the nearest neighbor density grids of adjacent slices (eg, for the $\Delta z = 0.3-0.4$ slice, we take the average standard deviation of the densities in $\Delta z = 0.2-0.3$ and $\Delta z = 0.4-0.5$), and normalized the density grid by this. The peak value was recorded as $\mathcal{S}$. For this method, $\mathcal{S}$ was determined from the nearest redshift slices that did not overlap the $\Delta z$ of the current slice.

After this broad search, we also conducted a narrower search with $\Delta z = 0.03 \cdot (1 + z_{prev})$ from $z=0$ to $z=4$ (at higher redshift there were too few objects per slice). This $\Delta z$ was selected to match the expected redshift error as determined from our accuracy tests, while also encapsulating the velocity range of a rich cluster, so each redshift slice should contain only objects with the potential to be closely associated. 

The overdensity candidates derived from both searches are summarized in Table 6, where we record any redshift slice for which $\mathcal{M} > 10$ and $\mathcal{S} > 10$ via either method, using as a cutoff the lowest-significance detection recorded by \citet{spi12}. We consulted the NASA/IPAC Extragalactic Database's list of known clusters to see if the search missed any known clusters. There weren't any listed clusters in the FIGS fields that were missed.

For comparison, we also ran the systematic search using photometric redshifts with $\Delta z = 0.03 \cdot (1 + z_{prev})$ from $z=0$ to $z=4$. The results of this comparison are summarized in Table 7. Generally, the photo-z search produced results of lower significance than the SPZ search. The photo-z search misses the $\mathcal{M},\mathcal{S} > 10$ cutoff for detection for several overdensities found by the SPZ method, and finds only two that the SPZ method doesn't detect (and one of these is marginal). Furthermore, the photo-z method finds the peak density for GN1-0.8 to be in the 0.870-0.926 redshift bin instead of 0.825-0.880, which we know from spectroscopic redshifts to be correct. This suggests that SPZs are better suited for accurately identifying known overdensities.

\subsection{A Potential Overdensity at $z=1.84$}

The known $z=0.85$ cluster in GN1 produced peaks of $\mathcal{M} = 25.16$ and $\mathcal{S} = 25.09$ with the broad search method. We find 4 other slices with a more significant detection in $\mathcal{S}$. Of these, the GS1/HUDF $\Delta z = 1.8-1.9$ slice is most significant in $\mathcal{M}$ with $\mathcal{M} = 18.20$. The density map for this slice is shown in Figure 13.

\begin{figure}
  \includegraphics[width=0.5\textwidth]{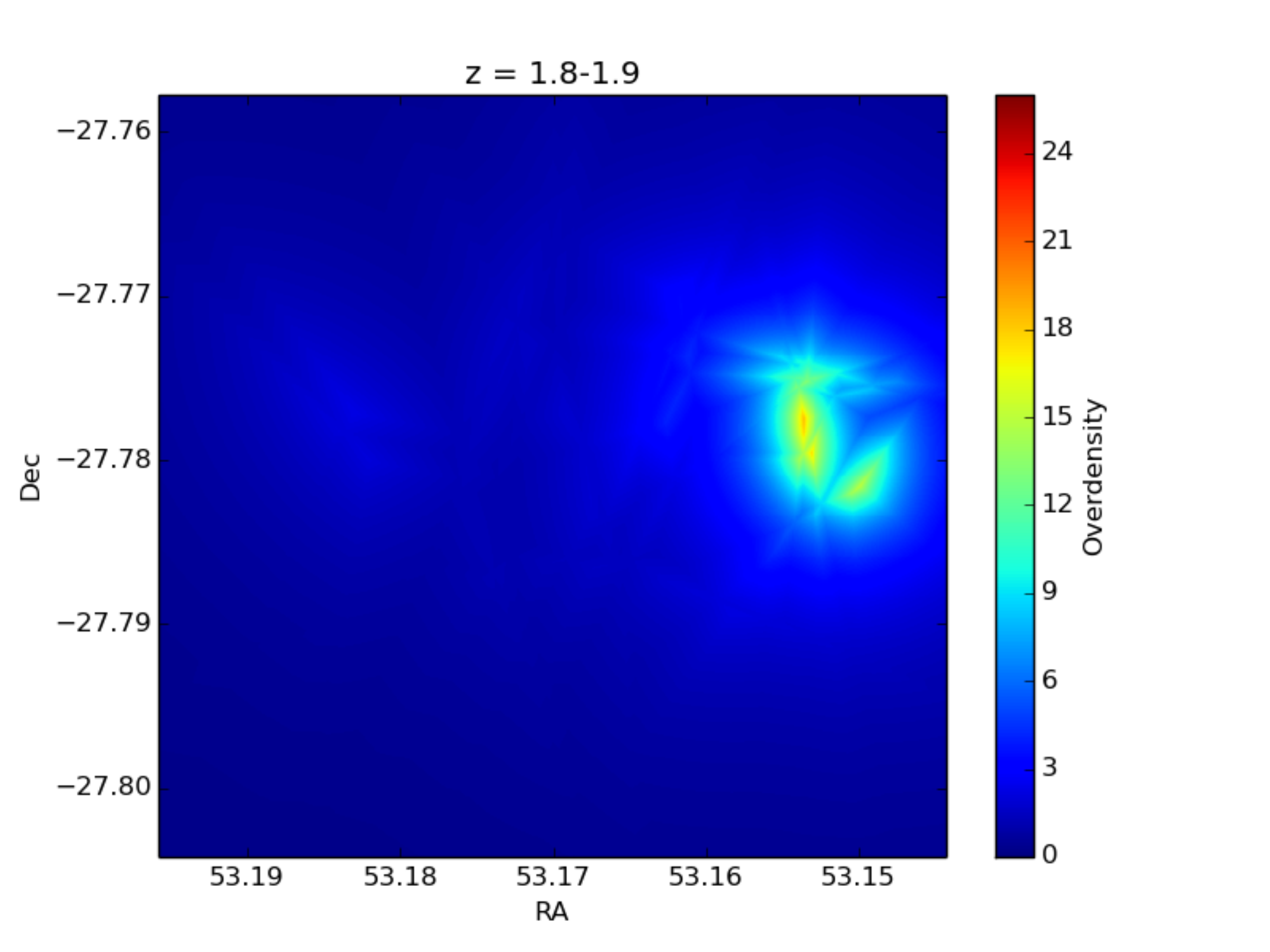}
  \caption{A 7th-nearest-neighbor density map for the GS1/HUDF redshift slice $\Delta z = 1.8-1.9$, normalized by the median density.}  
\end{figure}

\begin{table}
  \begin{center}
    \caption{SPZ Objects in $\Delta z = 1.8-1.9$}
    \begin{tabular}{cccccc}  
      \tableline
      FIGS ID & RA\footnotemark[1] & Dec & $z_{SPZ}$ & $z_{spec}$ & F105W \\
      \tableline
      1482 & 53.148895 & -27.777508 & 1.815 & 2.067 & 24.90 \\
      1601 & 53.157875 & -27.779194 & 1.815 & - & 26.02 \\
      3040 & 53.162968 & -27.800512 & 1.815 & - & 25.03 \\
      4300 & 53.15081  & -27.769133 & 1.815 & - & 25.47 \\
      \tableline
      4284 & 53.184544 & -27.768220 & 1.827 & - & 25.50 \\
      \tableline
      1049 & 53.172508 & -27.771004 & 1.843 & - & 25.83 \\
      1477 & 53.158291 & -27.777449 & 1.843 & - & 24.49 \\
      1623 & 53.154522 & -27.779718 & 1.843 & 1.837 & 24.03 \\
      1664 & 53.15287  & -27.780123 & 1.843 & - & 24.80 \\
      1781 & 53.149021 & -27.781952 & 1.843 & - & 24.13 \\
      \tableline
      1061 & 53.15604 & -27.770947 & 1.871 & - & 25.13 \\
      1524 & 53.148975 & -27.778151 & 1.871 & - & 26.05 \\     
      2091 & 53.192116 & -27.785559 & 1.871 & - & 26.37 \\
      4197 & 53.187511 & -27.76623  & 1.871 & - & 26.59 \\
      4258 & 53.152287 & -27.770088 & 1.871 & 1.852 & 23.66 \\
      4322 & 53.188129 & -27.768982 & 1.871 & - & 26.06 \\
      \tableline
      1499 & 53.152458 & -27.7777   & 1.9   & - & 25.83 \\
      1167 & 53.170788 & -27.772615 & 1.9  & - & 26.57 \\
      1905 & 53.182251 & -27.783314 & 1.9   & - & 24.93 \\
      2010 & 53.145897 & -27.784681 & 1.9   & - & 25.20 \\
      2266 & 53.192822 & -27.787857 & 1.9   & - & 27.35 \\
      4177 & 53.186508 & -27.768625 & 1.9   & - & 27.38 \\                  
      \tableline
    \end{tabular}
    \begin{tablenotes}
      \item[] $\rm ^a$ Using J2000.0 coordinates
      \end{tablenotes}
  \end{center}
\end{table}

\begin{table}
  \begin{center}
    \caption{SPZ Objects in $\Delta z = 0.075-0.107$}
    \begin{tabular}{cccccc}  
      \tableline
      FIGS ID & RA\footnotemark[1] & Dec & $z_{SPZ}$ & $z_{spec}$ & F105W \\
      \tableline
      1052 & 53.275558 & -27.859404 & 0.083 & - & 24.78 \\
      1164 & 53.278240 & -27.853859 & 0.083 & - & 24.20 \\
      1463 & 53.263836 & -27.866632 & 0.083 & - & 24.86 \\
      1491 & 53.280373 & -27.867067 & 0.083 & - & 25.28 \\
      3121 & 53.285057 & -27.841299 & 0.083 & - & 26.41 \\
      3303 & 53.284992 & -27.849686 & 0.083 & - & 24.16 \\
      3318 & 53.274483 & -27.850365 & 0.083 & - & 22.24 \\
      \tableline
      1042 & 53.276077 & -27.859423 & 0.094 & - & 22.10 \\
      1053 & 53.275337 & -27.859568 & 0.094 & - & 23.28 \\
      \tableline
      1098 & 53.278564 & -27.860065 & 0.105 & - & 26.48 \\
      1139 & 53.278801 & -27.860723 & 0.105 & - & 26.33 \\
      1156 & 53.278751 & -27.860992 & 0.105 & - & 26.19 \\
      1204 & 53.272770 & -27.861732 & 0.105 & - & 25.16 \\
      1316 & 53.286377 & -27.864700 & 0.105 & 0.1337 & 22.62 \\
      1364 & 53.258823 & -27.864935 & 0.105 & 0.1275 & 22.48 \\
      3439 & 53.277348 & -27.861378 & 0.105 & - & 25.99 \\              
      \tableline
    \end{tabular}
    \begin{tablenotes}
      \item[] $\rm ^a$ Using J2000.0 coordinates
      \end{tablenotes}
  \end{center}
\end{table}

The location of this overdensity matches that of a $z=1.84$ overdensity identified in \citet{mei15} through the visual inspection of G141 spectra and redshifts, and in \citet{koc15} by a search of NIR narrowband-selected emission-line galaxies. Mei et al. identifies 13 candidate members of a $z=1.84$ protocluster at (53.15565, -27.77930, J2000.0) at a limiting magnitude of F160W $< 26$ mag, as well as a number of nearby possibly associated galaxy groups at $z=1.87-1.95$. This is very near to the point of peak SPZ density (53.15357, -27.77756, J2000.0) identified via the nearest-neighbor method.

This redshift slice contains 22 objects with SPZs, for 3 of which we have matching spectroscopic redshifts. The characteristics of these objects are summarized in Table 8, where they are grouped by FIGS redshift. Two of the 3 are consistent with $\Delta z = 1.8-1.9$, and the third is $z=2.067$. Furthermore, in the CDFS 7 Ms X-ray source catalogs \citep{luo17}, we were able to visually identify a number of close X-ray active sources at this redshift, most of which are spectroscopically confirmed. The SPZ overdensity, when combined with the Mei et al. overdensity detection and possible presence of X-ray sources, suggests further corroboration of the use of SPZs to identify LSS via this method. Furthermore, it opens up the possibility of using SPZ searches to identify fainter candidate cluster members in already-identified overdensities at comparable redshift (eg, a GOODS-S but non-HUDF cluster identified in \citet{kur09}).

With the narrower search method, the $z=0.85$ cluster achieves similar significance, with $\mathcal{M} = 28.23$ and $\mathcal{S} = 24.41$. The $z=1.84$ overdensity in GS1 measures a lower but still significant detection via the $\mathcal{S}$ method, but measures a slightly higher significance via the $\mathcal{M}$ method. There are 4 other detections with a greater significance in at least one metric, including a hugely significant detection in GS2 at $\Delta z = 0.075-0.107$ (described in Table 9), and a similar number of detections at comparable significance. The GS2 detection does not appear to match any known overdensity, though it could potentially be associated with nearby diffuse X-rays at $z=0.126-0.128$ \citep{fin15}. This is also consistent with the two objects with known spectroscopic redshifts.

\section{Implications for Future Surveys}

The availability of grism spectra for computing SPZs via this method will only increase, as grism surveys will be a key component of future space missions, including the {\it James Webb Space Telescope} \citep{gar06, jan17} and the {\it Wide Field Infrared Survey Telescope}. For example, NIRISS on {\it JWST} will provide multi-object slitless spectroscopy at slightly lower resolution but with $\sim5$ times the wavelength coverage compared to HST/WFC3. {\it WFIRST} is anticipated to include a grism mode with $R \sim 550-880$ \citep{spe15} in the near-infrared. The expanded wavelength coverage of NIRISS and the increased resolution of {\it WFIRST}'s grism will allow surveys with either instrument to obtain redshifts via the method we describe with improved accuracy and outlier rates. This should result in a much larger collection of high-accuracy redshifts than are obtainable with ground-based spectroscopy alone. With FIGS, we produced $\sim1900$ redshifts for four 2.05'x2.27' fields, roughly three times the available number of spectroscopic redshifts and complete down to F105W $< 26.5$ mag. Wide-field slitless spectroscopy with NIRISS will operate with a similar 2.2'x2.2' field of view.

This will have major implications for cosmological studies conducted with the new instruments. The wide field of {\it WFIRST} and Euclid and the deep reach of {\it JWST} will enable more thorough LSS studies via cluster identification and weak lensing studies of a vast number of objects. Systematic LSS analyses will require dividing cluster and lensing samples into precise redshift slices, making redshift accuracy a measure of key importance. The application of grism spectra to the redshift measurement via this method can significantly expand the number of objects that are usable for such studies, enabling more and better cluster identifications and improved LSS science. With FIGS, we were able to identify a serendipitously identified and spectroscopically confirmed cluster without reliance on spectroscopic redshifts, which could provide a few advantages for LSS searches, as demonstrated via our systematic overdensity search. Furthermore, the accuracy of SPZs may allow for the identification of high-z overdensities in regions where spectroscopic redshifts are not plentiful. It is also possible that grism-enabled analysis of spectroscopically confirmed clusters can provide additional information about substructure within overdensities.


\section{Conclusions}

FIGS is a WFC3-G102 grism survey from which we obtained $\sim6000$ galaxy spectra, which we have combined with broadband photometry in order to produce more accurate spectrophotometric redshifts (called SPZs). Across all four fields and all magnitudes, we achieve a median $\Delta z /(1+z_{spec})$ of 0.02 for SPZs, as compared to 0.03 for pure photometric redshifts, uncorrected for the slight systematic bias described in \S 4. The SPZs also featured a lower rate of catastrophic failure in redshift fits (8\% to 7\% overall). SPZs provide an accurate redshift measurement for a larger number of objects per field than can be achieved with ground-based spectroscopy. As grism surveys become more common in upcoming missions, this will allow for the calculation of more comprehensive catalogs of high-accuracy redshifts.

Analysis of the redshift distributions in the SPZs enabled us to independently identify a previously spectroscopically confirmed galaxy cluster at $z=0.85$, and to identify a known overdensity at $z=1.84$ using the nearest-neighbor density method. Applying this method systematically across redshift slices in the FIGS fields, we were also able to detect a potentially new overdensity at $z \sim 0.1$, and four other candidate overdensities with a significance comparable to that of the $z=0.85$ cluster in at least one measure. Given the higher accuracy of SPZs compared to photometric redshifts, this suggests an alternative to detect large scale structure in regions where spectroscopic redshifts are rare. SPZs can also provide the identification of additional cluster member galaxies, which may make it possible to better analyze substructure within a cluster. 

\section{Acknowledgements}

We thank the referee for their helpful comments, which have greatly improved this paper. We also thank Professor Lori Lubin of the UC Davis Physics Department for her comments and insight on galaxy clusters.

Based on observations made with the NASA/ESA Hubble Space Telescope, obtained [from the Data Archive] at the Space Telescope Science Institute, which is operated by the Association of Universities for Research in Astronomy, Inc., under NASA contract NAS 5-26555. These observations are associated with program \#13779.

Support for program \#13779 was provided by NASA through a grant from the Space Telescope Science Institute, which is operated by the Association of Universities for Research in Astronomy, Inc., under NASA contract NAS 5-26555.

AC acknowledges the grants ASI n.I/023/12/0 "Attività relative alla fase B2/C per la missione Euclid" and MIUR PRIN 2015 "Cosmology and Fundamental Physics: illuminating the Dark Universe with Euclid".

{}

\end{document}